\documentclass[prd,twocolumn,showpacs,preprintnumbers,nofootinbib,amsmath,amssymb,superscriptaddress,nob
alancelastpage]{revtex4}

\usepackage{graphicx}
\usepackage{dcolumn}
\usepackage{bm}

\newcommand{\beq}{\begin{equation}}
\newcommand{\eeq}{\end{equation}}

\newcommand{\Mmax}{M_{\mathrm{max}}}
\newcommand{\Mmin}{M_{\mathrm{min}}}

\newcommand{\rt}{r_t}

\begin{document}

\title{The distribution of annihilation luminosities in dark matter substructure}

\author{Savvas M. Koushiappas}
\affiliation{Department of Physics, Brown University, 182 Hope Street, Providence, RI 02912}
\email{koushiappas@brown.edu}
\author{Andrew R. Zentner}
\affiliation{Department of Physics \& Astronomy, University of Pittsburgh, Pittsburgh, PA 15260}
\email{zentner@pitt.edu}
\author{Andrey V. Kravtsov}
\affiliation{Kavli Institute for Cosmological Physics, The University of Chicago, Chicago, IL 60637}
\affiliation{Enrico Fermi Institute, The University of Chicago, Chicago, IL 60637}
\affiliation{Department of Astronomy \& Astrophysics, The University of Chicago, Chicago, IL 60605} 
\email{andrey@oddjob.uchicago.edu}


\begin{abstract}
We calculate the probability distribution function (PDF) of the
expected annihilation luminosities of dark matter subhalos as a
function of subhalo mass and distance from the Galactic center using a
semi-analytical model of halo evolution. We find that the PDF of
luminosities is relatively broad, exhibiting a spread of as much as an
order of magnitude at fixed subhalo mass and halo-centric distance. 
The luminosity PDF allows for simple
construction of mock samples of $\gamma$-ray luminous subhalos and
assessment of the variance in among predicted $\gamma$-ray signals from 
dark matter annihilation.  Other applications include quantifying the variance among the
expected luminosities of dwarf spheroidal galaxies, 
assessing the level at which dark matter annihilation can be a
contaminant in the expected $\gamma$-ray signal from other
astrophysical sources, as well as estimating the level at which nearby 
subhalos can contribute to the antimatter flux.
\end{abstract}

\maketitle

\section{Introduction}
\label{sec:introduction}

Multiple lines of observational evidence have established the
existence of a form of non-baryonic dark matter binding galaxies.
Dark matter is commonly considered a yet-to-be discovered elementary
particle.  A promising candidate is a Weakly-Interacting Massive
Particle (WIMP) that arises in extensions to the standard model of
particle physics.  Examples include the lightest supersymmetric
particle \cite{Jungman:1995df,BHS05} and particle excitations in
theories of Universal Extra Dimensions \cite{Hooper:2007qk}.  WIMPs
interact via the Weak interaction and were in thermal equilibrium in
the early Universe.  If the dark matter is a thermal relic WIMP, the
WIMP annihilation cross section can be constrained by requiring that
the present dark matter density is $\Omega_{\mathrm{c}} \approx 0.23$
\cite{Komatsu:2008hk}.  Implied values for this cross section are of
order $\langle \sigma v \rangle \sim 10^{-26}$~cm$^3$/s, and candidate
WIMPs can annihilate to many observable states, such as $\gamma$-rays
and high-energy neutrinos.

The Cold Dark Matter (CDM) model of cosmological structure formation
predicts that dark matter is distributed in ``halos" in a hierarchical
fashion.  The host halo of the Milky Way is expected to have mass of
$M \approx 10^{12}\ h^{-1}$M$_{\odot}$
\cite{Klypin:2001xu,Li:2007eg}. Within the CDM model such halos are also
expected to contain numerous smaller dark matter ``subhalos" which, in
turn, contain subhalos of their own, perhaps with masses all the way
down to the cutoff scale in the primordial density fluctuation power
spectrum \cite{Schmid:1998mx,HSS01,Green:2003un,Green:2005fa}.

A well-studied avenue for possible identification of the dark matter
is to search for unique products of dark matter annihilation at the
center of our Galactic halo, where densities are highest
\cite{Dodelson:2007gd,Serpico:2008ga,Gondolo:1999ef,
Horns:2004bk,Merritt:2002vj,Bergstrom:1997fj,Aloisio:2004hy,Zaharijas:2006qb}.
However, astrophysical backgrounds are also highest toward the
Galactic center, so an alternative approach is to search for
annihilation within subhalos
\cite{Baltz:1999ra,Tyler:2002ux,Evans:2003sc,Profumo:2005xd,Bergstrom:2005qk,Calcaneo-Roldan:2000yt,Tasitsiomi:2002vh,Stoehr:2003hf,Koushiappas:2003bn,
Baltz:2006sv,Strigari:2006rd,Strigari:2007at,Diemand:2006ik}
(including potentially sub-solar mass halos
\cite{Bringmann:2009vf,Koushiappas:2009du,Bringmann:2006mu,Pieri:2005pg,Ando:2008br,Diemand:2006ik}),
or to infer annihilation within subhalos statistically through the
angular distribution of diffuse $\gamma$-ray emission
\cite{Ando:2009fp,Ando:2006cr,Cuoco:2006tr,Cuoco:2007sh,
Fornasa:2009qh,Hooper:2007be,Lee:2008fm,SiegalGaskins:2009pz,
SiegalGaskins:2009ux,SiegalGaskins:2008ge,Taoso:2008qz}.  More
recently, the one-point $\gamma$-ray flux probability distribution
function (PDF) was proposed as another way to quantify the expected
annihilation signal \cite{Lee:2008fm} .  The main idea behind this
approach is that the pixel-to-pixel flux variation from $\gamma$-ray
emission from small dark matter subhalos would deviate from the Poisson
fluctuations that would be expected from a smoothly-distributed dark
matter halo and smoothly-distributed backgrounds.

A calculation of the contribution of annihilation products from
subhalos to the flux along any sight line consists of the following
two ingredients.  The first is the number of subhalos intercepted
along the line of sight, assuming subhalos to be small compared to the
angular resolution of the instrument as is the case with contemporary
detectors.  The second is the annihilation luminosity of each
intercepted subhalo.  The latter depends on the distribution of dark
matter within subhalos, which reflects the underlying process of
nonlinear mass assembly individual to each subhalo.  Although two
subhalos may have the same mass and be located at the same
Galacto-centric distance today, they may have different annihilation
luminosities because they may have different formation times, mass
assembly histories, and different orbits in the Milky Way potential.
Each of these factors affects the internal densities of subhalos.
Therefore, we expect a distribution of luminosities at each subhalo
mass and Galacto-centric distance.

In this paper, we estimate the probability distribution function (PDF)
of the subhalo luminosity as a function of the subhalo mass and
Galacto-centric distance from a large ensemble of subhalo populations
generated using a semi-analytic model of halo and subhalo evolution
\cite{Zentner:2004dq}.  Our aim is to use a statistically-large sample
of subhalo properties to provide a useful tool for computing subhalo
annihilation signals in a way that captures some of the complexity of
nonlinear evolution.  We show that the luminosity PDF of subhalos is
well fit by a log-normal distribution, reflecting the underlying
distribution in formation times (or concentrations,
\cite{Bullock:1999he,Wechsler:2001cs}) and provide simple, empirical
fits to the luminosity PDFs as a function of mass and distance.
Applications for the derived PDF range from quantifying the variance
in the expected luminosities of dwarf spheroidal galaxies to
generating mock synthetic $\gamma$-ray sky maps, to understanding the
level at which dark matter annihilation can be a contaminant in the
expected $\gamma$-ray signal from other astrophysical sources
\cite{Ando:2006mt,Miniati:2007ke,Zhang:2004tj,Pavlidou:2002va}, and to 
ascertain the level at which a nearby subhalo can contribute to the 
measured flux of antimatter \cite{Hooper:2008kv}.

\section{The importance of the subhalo luminosity PDF} 
\label{section:importance}
 
The effects of a distribution of luminosities on any calculation that
involves the diffuse emission is only important if most of the diffuse
flux originates from a large number of sources.  If, for example, the
diffuse flux is due to very few sources with high luminosities, then
the properties of the diffuse background would be dominated by the
Poisson statistics of the emission of photons from these sources.  If,
on the other hand, there are many dim sources along the line of sight,
the intrinsic variation in the luminosities of these sources will have
an effect on the flux PDF. This is due to the fact that the flux PDF 
will deviate from Poisson statistics as it will depend not only on the 
flux-density distribution but also on the mean number of sources (see
the '$P(D)$ analysis' discussion in the Appendix of \citet{Lee:2008fm}).

\begin{figure*}[t]
\includegraphics[height=7cm]{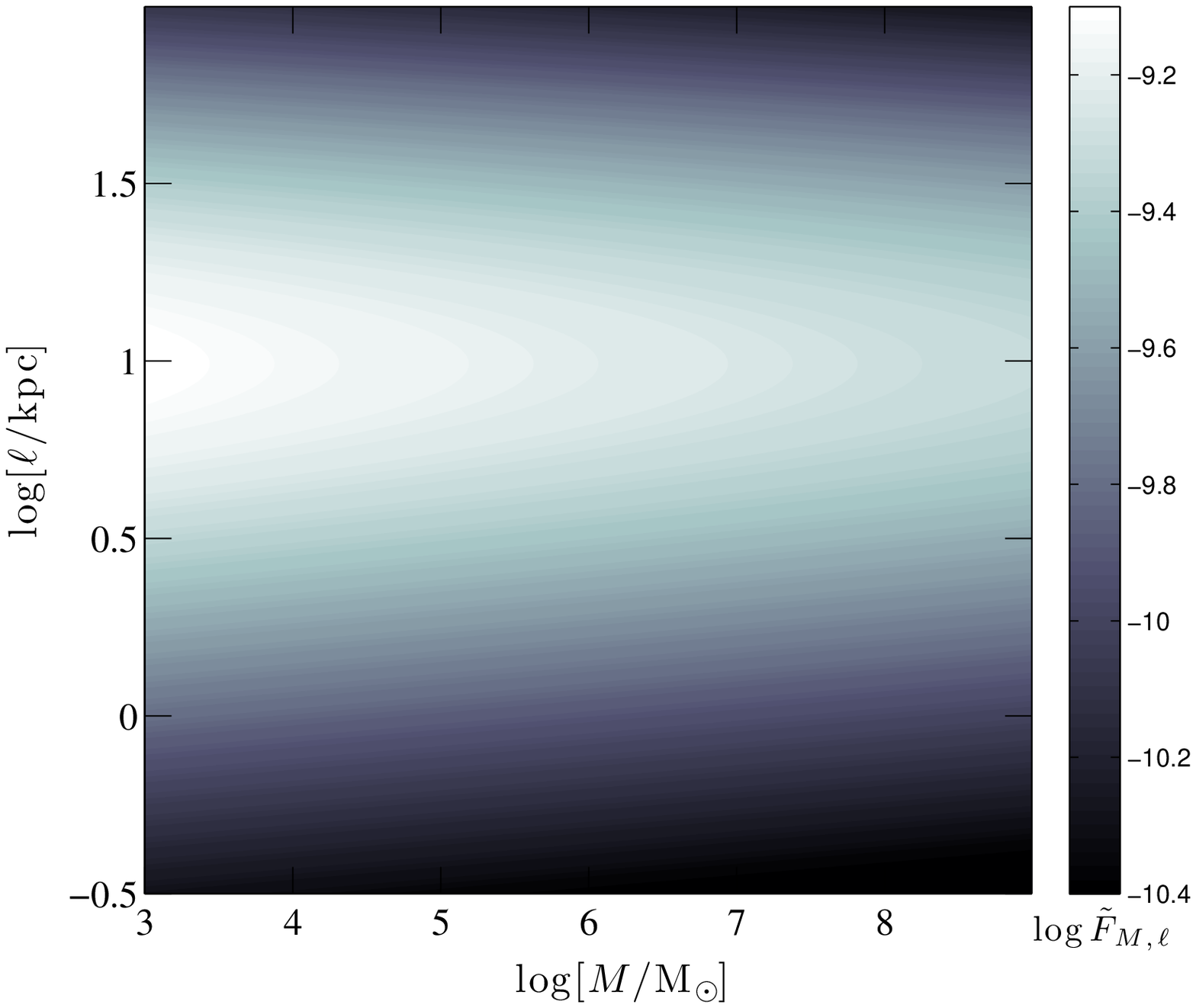}
\includegraphics[height=7cm]{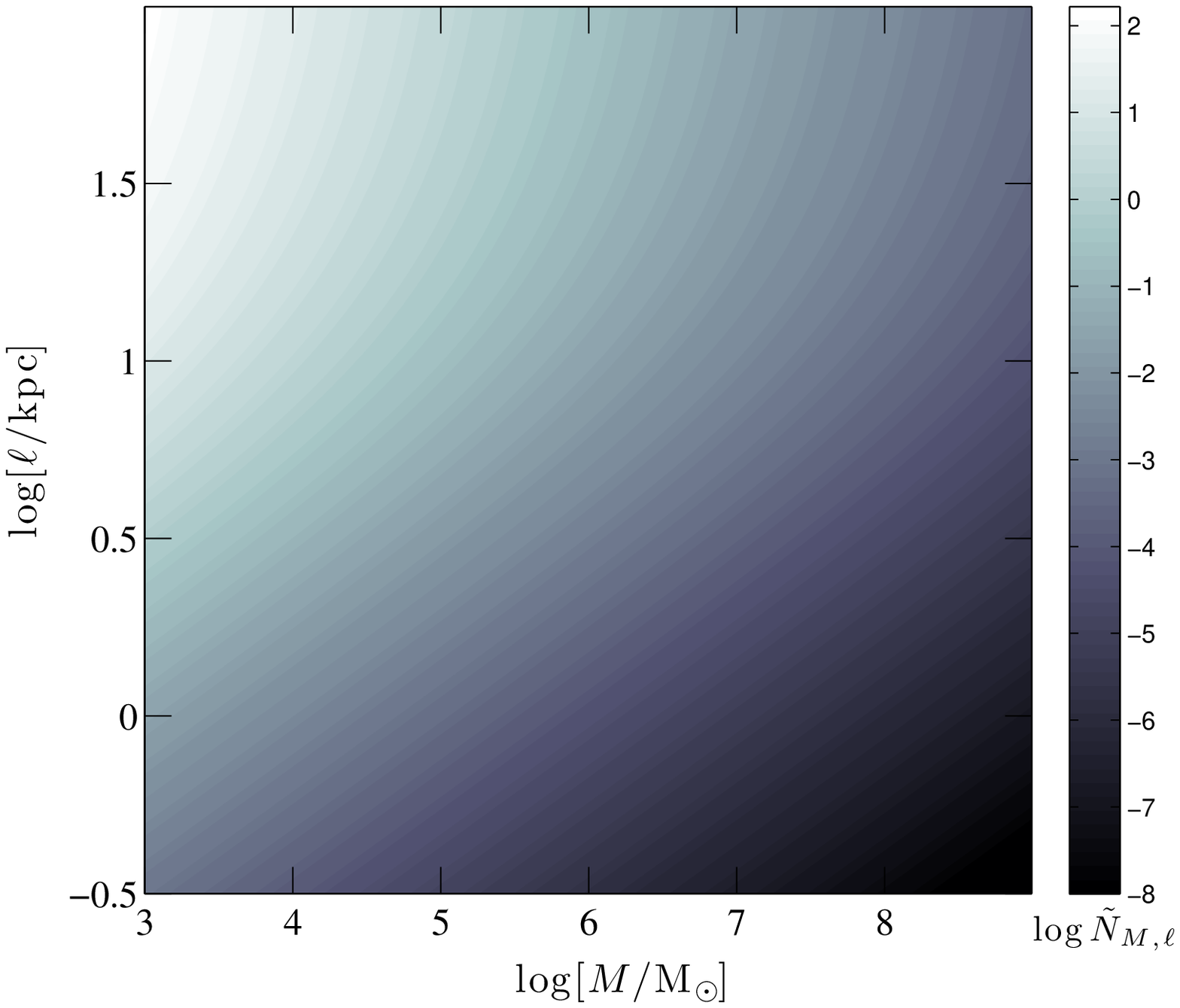}
\caption{
{\it Left}:  The mean contribution to the annihilation flux from all subhalos 
in a logarithmic interval of subhalo mass (horizontal axis) and 
line-of-sight distance (vertical axis), $\tilde{F}_{M,\ell}$ defined 
in Eq.~(\ref{eq:Ftilde}) in units of ${\mathrm{cm}}^{-2}{\mathrm{s}}^{-1} {\mathrm{sr}}^{-1}$.  This panel shows the mean contributions along a 
line of sight $90^{\circ}$ away from the Galactic center.  
The color at each position designates the 
value of $\tilde{F}_{M,\ell}$, where lighter colors indicate larger 
values as shown in the color bar.   
The contribution along the line of sight peaks at $\sim 10$~kpc at all masses.  
{\it Right}: The mean number of subhalos per logarithmic interval of subhalo 
mass and logarithmic line-of-sight interval.  Lighter colors designate 
larger numbers of subhalos according to the color bar.  Smaller objects
dominate the number density at all radii [see Eq.~(\ref{eq:Ntilde})].
}
\label{fig:fig0}
\end{figure*} 

We can quantify this argument as follows.  Suppose the number density
of subhalos of luminosity $L$ at position $\ell$ along a particular
line of sight is given by $dN/dMdV \propto M^{-\alpha} n[\ell(r)]$,
where $n[\ell(r)]$ is the number density of subhalos at
Galacto-centric distance $r$ along line of sight distance $\ell$.
Additionally, assume a mapping between the luminosity and the mass of
a subhalo, $L\propto M^\beta$.  Then the contribution to the received
flux that is produced by subhalos in a given logarithmic mass and line
of sight interval is given by
\begin{equation} 
\tilde{F}_{M,\ell} \equiv \frac{d F}{d \ln M d \ln \ell}  \propto  M^{-\alpha + \beta + 1} \ell \, n[r(\ell)],  
\label{eq:Ftilde}
\end{equation}
Taking $\alpha \approx 1.9$ and $\beta \approx 0.8$ (consistent with
analytical arguments \cite{Strigari:2006rd}, and numerical simulations
\cite{Stoehr:2003hf,Diemand:2006ik,kuhlen:2008aw}), the flux per
logarithmic interval in mass and line of sight distance has a weak
dependence on mass $\tilde{F}_{M,\ell} \propto M^{-0.1}$. 

In order for the low mass subhalo contribution to the annihilation
flux to be roughly the same as high mass halos, their abundance must
be larger.  The mean number of subhalos per logarithmic line of sight
interval and logarithmic mass interval is
\begin{equation}
\tilde{N}_{M,\ell} \equiv  \frac{ dN}{d \ln M d \ln \ell}  \propto  M^{-\alpha + 1} \ell^3 \, n[r(\ell)]. 
\label{eq:Ntilde}
\end{equation}
Assuming that the distribution of subhalos traces that of dark matter,
$n[\tilde{r}(\ell)] \propto [\tilde{r}( 1 + \tilde{r})]^{-1}$
\cite{NFW96,Navarro:1996gj}, where $\tilde{r} = r/r_s$ and $r_s$ is the scale radius
of the NFW dark matter profile.  For $\alpha \approx 1.9$ and small
distances ($\ell \ll r \ll r_s$), $n[\tilde{r}(\ell)] \propto
1/\tilde{r}(\ell)$ so that $\tilde{N}_{M,\ell} \propto M^{-0.9}
\ell^2$.  For large distances ($\ell \approx r \gg r_s$),
$n[\tilde{r}(\ell)] \propto 1/\tilde{r}^3(\ell)$ and
$\tilde{N}_{M,\ell} \propto M^{-0.9}$, roughly independent of $\ell$
for all masses.

The previous two paragraphs show that we should expect the majority of
the diffuse annihilation flux to be due to the presence of numerous
low-mass subhalos.  In the left panel of Fig.~\ref{fig:fig0} we show
the quantity $\tilde{F}_{M,\ell}$ in units of ${\rm cm}^{-2}{\rm
s}^{-1}{\rm sr}^{-1}$, as a function of subhalo mass and position
along the line of sight, Eq.~(\ref{eq:Ftilde}), at an angle $\psi =
90^\circ$ with respect to the Galactic center.  We assume a Milky Way
halo with a radius $R_{MW} = 250 \, {\rm kpc}$, and a scale radius
$r_s \approx 20 \, {\rm kpc}$.  Note that at any fixed mass, most of
the flux comes from a region at $\ell \approx 10 \, {\rm kpc}$.  This
is because the number of objects declines rapidly with distance at
Galacto-centric distances signficantly larger than a scale radius.
Moreover, low-mass subhalos contribute marginally more flux than their
high-mass counterparts.  In the right panel of Fig.~\ref{fig:fig0} we
show the mean number per logarithmic mass and line-of-sight interval,
$\tilde{N}_{M,\ell}$.  The two panels of Fig.~\ref{fig:fig0}
illustrate a basic conclusion that the mean number of objects along a
line of sight increases with decreasing subhalo mass (right panel),
and as all intervals of subhalo mass have comparable flux
contributions (left panel), the signal is set by relatively low-mass
subhalos that are close to the observer.  This is in qualitative
agreement with the result of \citet{Lee:2008fm} who found that
substructures give rise to photon counts that deviate from a Poisson
distribution.

Our simple demonstration neglects the presence of a baryonic disk in
the Milky Way halo, and its effect on the subhalo population in the
inner regions of the halo.  Recent studies find that the inner 30~kpc
of a Milky Way-sized halo may be deficient in subhalos due to
interactions with the disk \cite{D'Onghia:2009pz}.  This should have
an effect on the flux contributions we derived, but the approximate
mass and distance dependence that we describe should be maintained.  
Nevertheless, it is important to keep in mind that a depleted subhalo
population in the inner regions of the Galactic disk may have an
effect on the flux from substructure.  The maximal net suppression in
annihilation flux near the disk due to this suppression is expected to
be a factor of a few \cite{D'Onghia:2009pz}.


\section{The subhalo annihilation luminosity PDF}
\label{section:pdf_methods}

In order to derive the subhalo luminosity PDF, we model the accretion
and dynamical evolution of subhalos in the Milky Way potential well
using an approximate semi-analytic technique \cite{Zentner:2004dq}.
This approximate approach greatly reduces the computational cost of
modeling substructure by treating subhalo density profiles as
continuously-evolving functions that can be described by a small
number of parameters, rather than a collection of a very large number
of individual particles.  This enables calculations of the properties
of subhalo populations in a large set of Milky Way-sized halos, so
that object-to-object variance can be estimated.  This is not yet
feasible in simulations directly.  This procedure also provides a
physically-motivated method to extrapolate the results of numerical
simulations to masses below their resolution limits, masses which are
not negligible from the annihilation perspective.  Of course, the cost
of this method is that it is approximate and the approximations
implemented cannot be validated outside the range of scales that are
resolved by $N$-body simulations.  This method is described in detail
in \cite{Zentner:2004dq}, which also shows a number of non-trivial
comparisons with $N$-body simulations that validate this treatment of
halo substructure.

We use an ensemble of 200 realizations of subhalo populations within a
dark matter halo of mass $M_{\rm MW} = 1.26 \times 10^{12} \,
h^{-1}$M$_\odot$ in a flat cosmological model with $\Omega_{dm} =
0.228$, $\Omega_b h^2 = 0.0227$, $h=0.71$, and $\sigma_8=0.81$,
favored by the five-year Wilkinson Microwave Anisotropy Probe results
\cite{Komatsu:2008hk}.  Each realization represents a possible subhalo
population within a Milky Way-like halo.  The populations differ
because only some statistical properties of the initial density field
in the local neighborhood are known, not the precise initial
conditions for collapse.  We use this large number of distinct
populations to quantify the predicted variation from one Milky
Way-sized halo to another.  The result of the calculation is a list of
all subhalos, complete with their structural parameters such as bound
mass, scale radius, and tidal radius.  These quantities all evolve
with time, and we study them at the present epoch ($z=0$).  

\begin{figure}
\includegraphics[height=8cm]{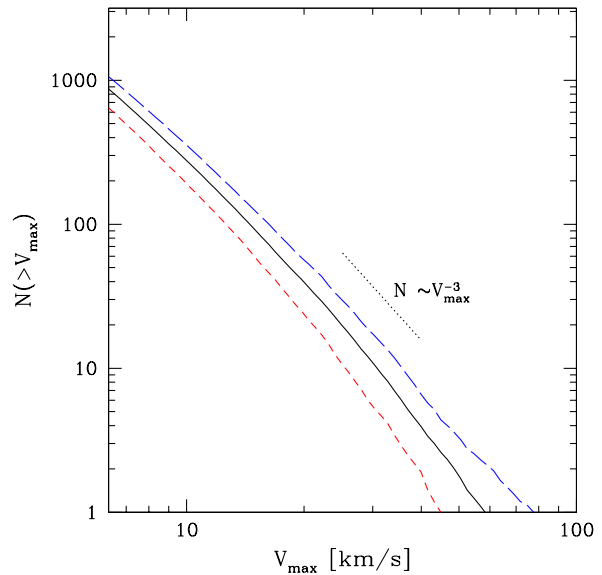}
\caption {The cumulative velocity function of subhalos. The solid
black line shows the average velocity function over all 200
realizations of the formation of a Milky Way-size halo. The
short-dashed red line shows the velocity function derived for Galactic
halos which have a concentration $c > 13.4$, while the long-dashed
blue line shows the velocity function for Galactic halos with $c <
6.7$.  The dotted line shows the behavior of a $N(>V_{\rm max}) \propto
V_{\rm max}^{-3}$ power law, which describes the subhalo velocity
function in cosmological simulations.}
\label{fig:fig1}
\end{figure}

Figure~\ref{fig:fig1} shows the cumulative velocity function of
Galactic subhalos. The power-law behavior is similar to what is found
in numerical simulations, with the number of subhalos increasing as
$N(>V_{\rm max}) \propto V_{\rm max}^{-3}$. Our model also reproduces
the abundance of subhalos in cosmological simulations reasonably well (see also
Ref.~\cite{Zentner:2004dq} for a discussion).  For example, the
average number of subhalos with $v_{\mathrm{max}} > 4$~km/s within the
inner 200 kpc of the Milky Way in our models is 2481 with a 68
percentile range of [1964-3007].  This is consistent with 2469
subhalos of $v_{\mathrm{max}} > 4$~km/s within the same radius in the
Via Lactea II simulation of a Milky Way-sized halo
\cite{Kuhlen:2008qj}.  Moreover, the dispersion in the number of
subhalos is consistent with the empirical finding of a Poisson scatter
added in quadrature to an intrinsic scatter of about $20\%$ derived
from recent numerical simulations \cite{BoylanKolchin:2009an}.  We note that the validation
exercises in Ref.~\cite{Zentner:2004dq} show that this model may
over-estimate halo-to-halo variance mildly (see their Fig.~7), though
it is difficult to assess through a comparison with simulations 
with greater precision at this point.

Note that in general, the slope and normalization of the velocity
function depend on the concentration of the host dark matter halo
\cite{Zentner:2004dq}.  As shown in Fig.~\ref{fig:fig1}, the
normalization of the velocity function of subhalos is reduced. This is
to be expected as host dark matter halos with high concentrations are
formed earlier than low mass halos.  As a result, there is more
available time for evolution of the subhalo population leading to a
decrease in the total number of subhalos. A subsidiary effect that
also contributes to the decrease of the subhalo population of
high-concentration halos is the efficient tidal disruption due to the
higher central densities.  On the other hand, subhalos in host halos
with low concentrations will, on average, have spent less time in the host
and have a higher rate of survival due to the decreased tidal
forces.

We model the final dark matter distribution in a subhalo of mass
$M$ as a Navarro-Frenk-White (NFW) profile \cite{NFW96,Navarro:1996gj},
\begin{equation}
\label{eq:nfw} \rho(r) = \frac{M}{4 \pi r_s^3}
\frac{1}{f(\tilde{r}_t)} \frac{1}{\tilde{r} ( 1 + \tilde{r} )^2}.
\end{equation}
In Eq.~(\ref{eq:nfw}), $\tilde{r} = r / r_s$, where $r_s$ is the
evolved scale radius.  $f(x) = \ln ( 1 + x ) - x / ( 1 + x )$, and
$\tilde{r}_t = r_t / r_s$ is the ratio of a tidal truncation radius
$r_t$, to the scale radius.  We treat tidal truncation by assuming an
abrupt limit to the extent of the subhalo density profile for
simplicity.  Isolated high-resolution simulations of subhalo evolution
support a very sharp truncation (e.g., Ref.~\cite{Kazantzidis:2003hb,Kazantzidis:2009zq}) and,
furthermore, the annihilation luminosity emanating from the transition
region should be very small in comparison to the total annihilation
luminosity. Our simplifying assumption of a truncated NFW profile is not 
crucial to our analysis.  Deviations in the assumed power law in the 
inner regions of the profile do not result in large changes in the total 
luminosity of the halo.  Profiles with different inner slopes also require 
different normalizations to maintain the same bound mass against the 
tidal forces of the host.  Moreover, much of the total luminosity of a 
nearly-NFW halo arises from the region near the scale radius, $r_s$.  
As such, changes in the inner slope do not lead to large changes in the 
annihilation luminosity in most circumstances 
(see Refs.~\cite{Koushiappas:2003bn,Bergstrom:2005qk,Strigari:2006rd,Robertson:2009bh,Reed:2010}).

The dark matter luminosity of a subhalo is then obtained from
\begin{eqnarray}
\label{eq:luminosity} L &=& 4 \pi \frac{ \langle \sigma v \rangle \,
N_{\gamma}^{\rm tot}}{M_{\chi}^2} \int_0^{\rt} \rho^2 r^2 dr \nonumber \\ 
& = & \frac{3.32 \times 10^{37}\, \mathrm{ph}}{\mathrm{s}} \  \frac{ \langle \sigma v
\rangle_{-26} \, N_{\gamma,30}^{\rm tot}}{M_{\chi,100}^2} \nonumber \\
&\times& \left(\frac{r_s}{{\mathrm{kpc}}} \right)^3
\int_0^{\tilde{r}_t} \left( \frac{
\rho(\tilde{r})}{{\mathrm{GeV/cm}}^{3}} \right)^2 \tilde{r}^2 d
\tilde{r} .
\end{eqnarray}
Here, $\langle \sigma v \rangle_{-26}$ is the annihilation cross
section in units of $3 \times 10^{-26} {\rm cm}^3 {\rm s}^{-1}$,
$M_{\chi,100}$ is the mass of the dark matter particle in units of 100
GeV, and $N_{\gamma,30}^{\rm tot}$ is the total number of photons
emitted above a threshold of 1 GeV, in units of 30.  This fiducial
choice of parameters is representative of optimistic scenarios in the
Minimal Supersymmetric Model.  We emphasize that we defined luminosity
as {\em number of photons per unit time}, and not energy per time.
Similarly, when we discuss flux, we implicitly mean photon flux, and
not the energy flux.  Note also that as the goal of this work is to
quantify the spread in luminosities in the subhalo population, the
choice of particle physics parameters in Eq.~\ref{eq:luminosity} is
intended only to provide a useful representation of the magnitude of
the luminosity. However, the PDFs of luminosities derived below are
affected by the choice of particle physics parameters only in their
normalization.

\section{Results}
\label{section:pdf_results}
%
\begin{figure*}[t!]
\includegraphics[height=5.9cm]{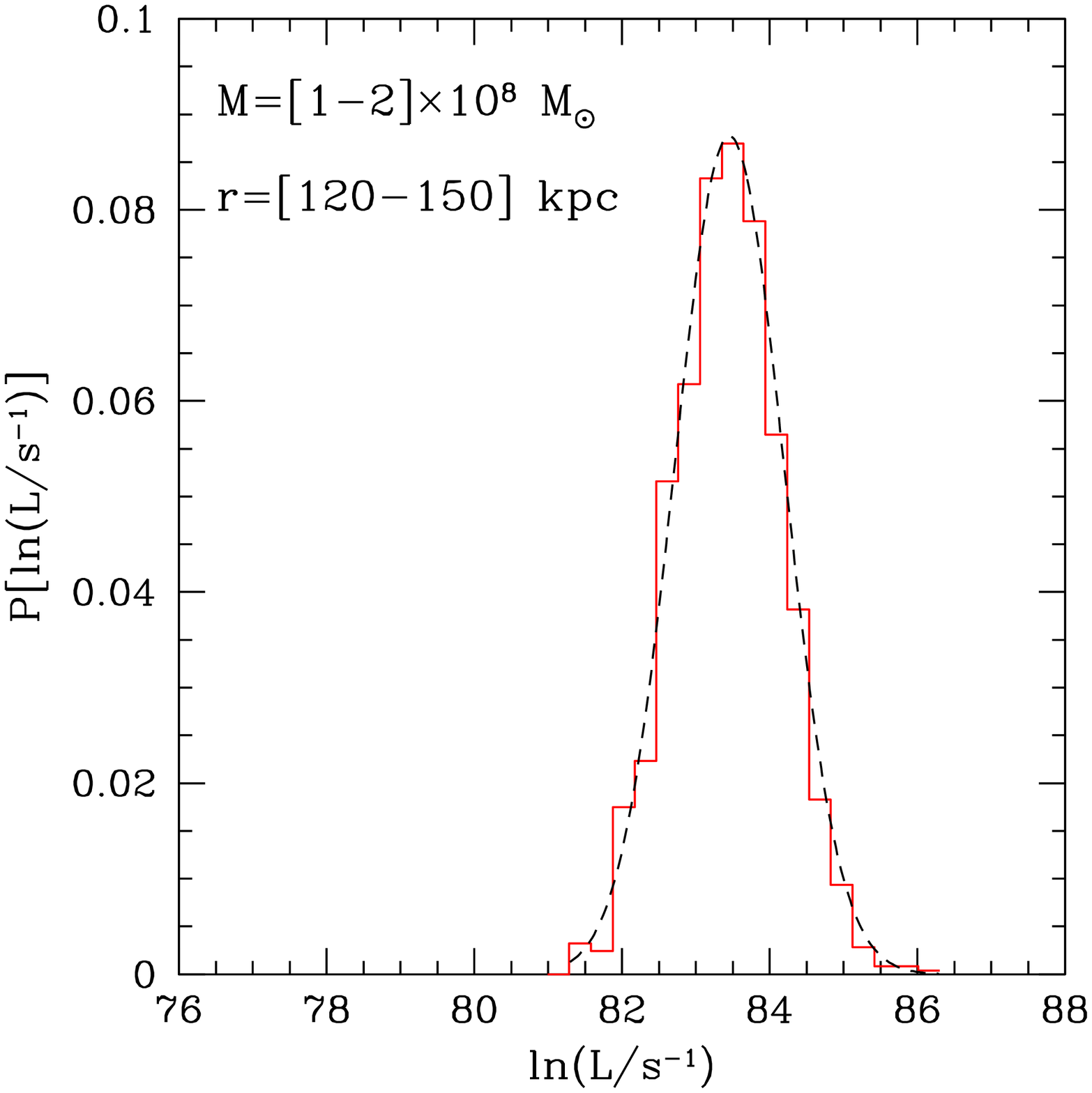}
\includegraphics[height=5.9cm]{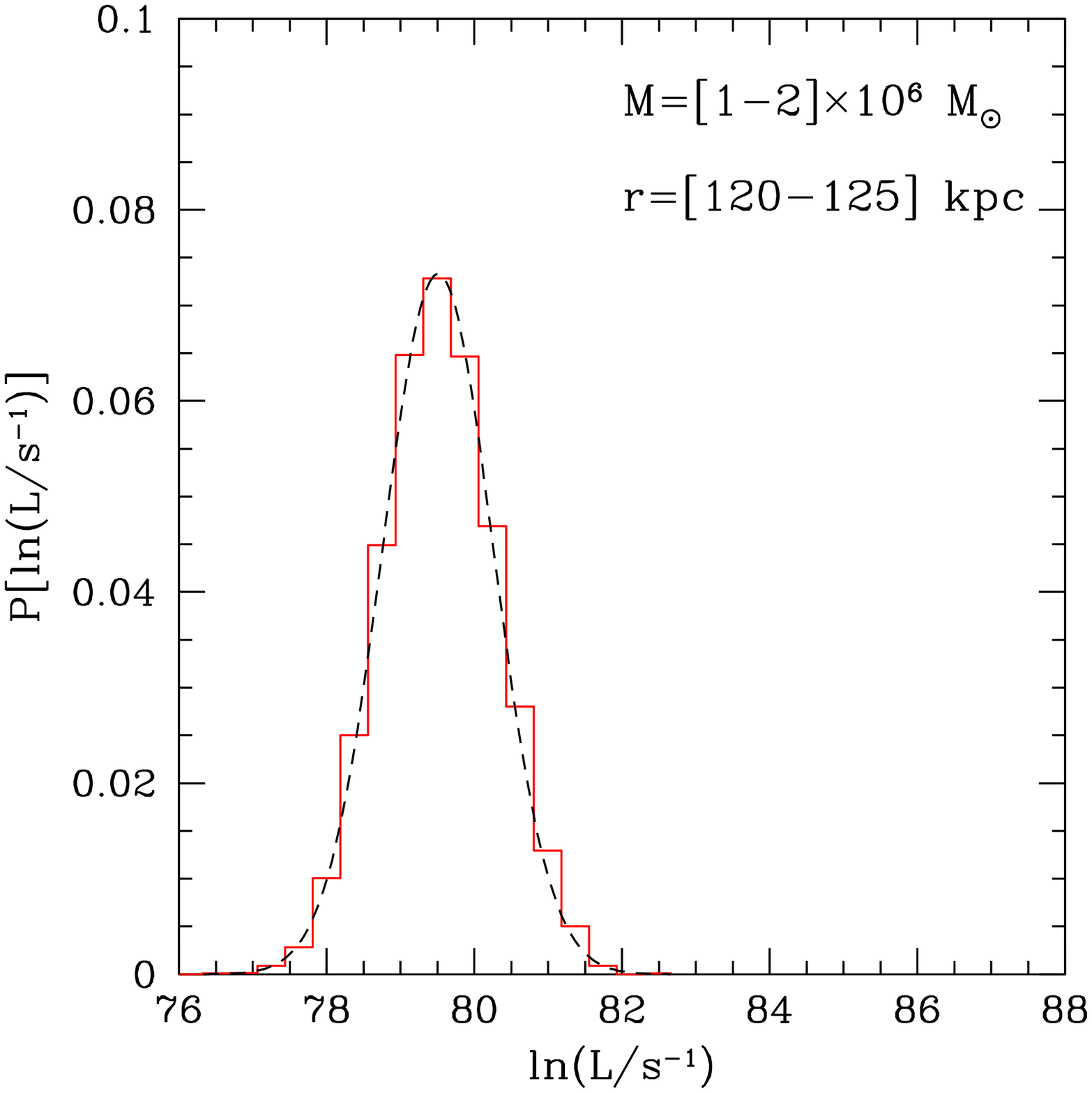}
\includegraphics[height=5.9cm]{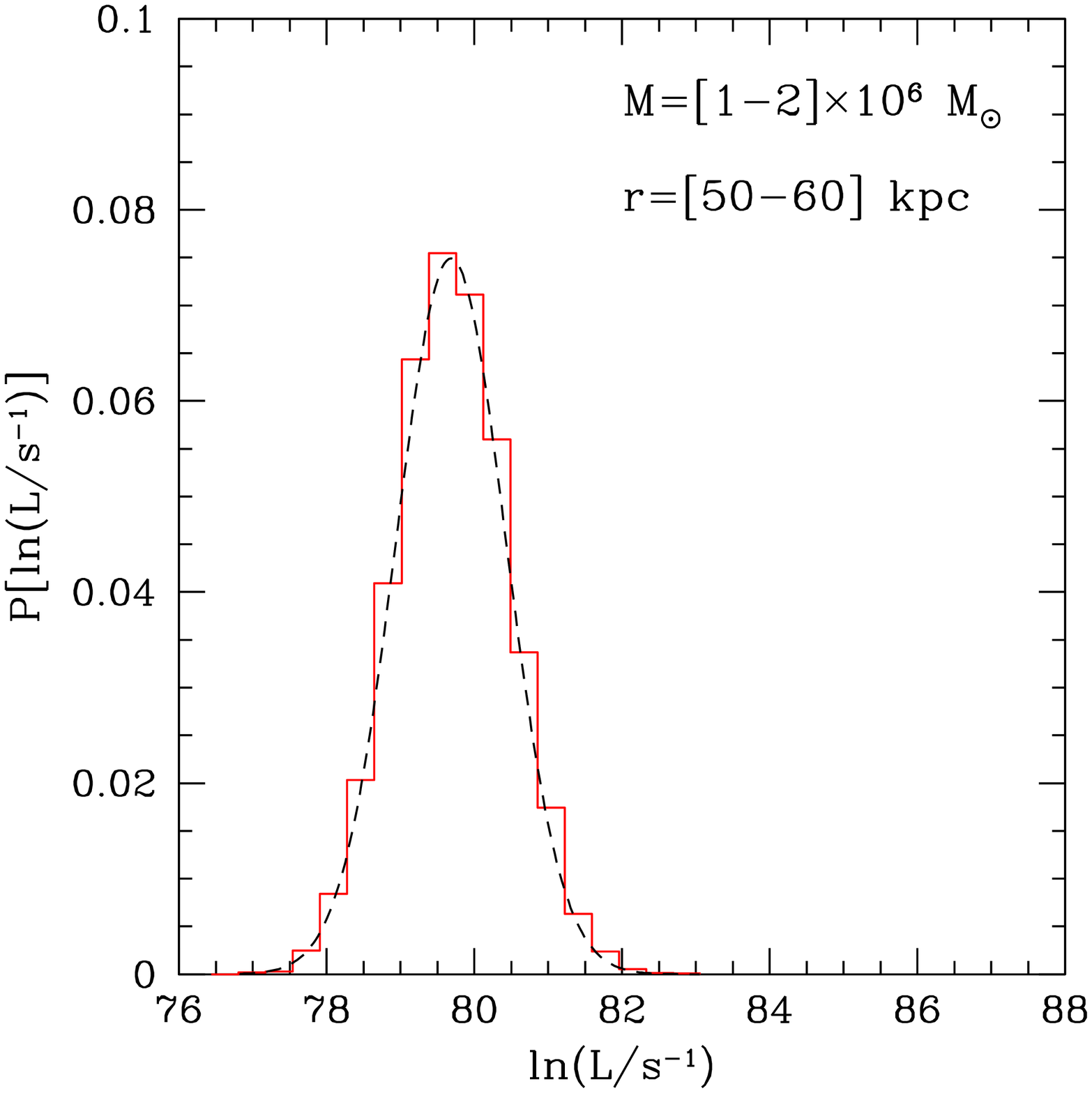}
\caption {Example probability distribution functions of the $\gamma$-ray
luminosity for three sample mass and radial bins. The solid curve
depicts a histogram of the distribution of luminosities in each
bin. The dashed curve shows the fitting function of
Eq.~\ref{eq:PofL}.}
\label{fig:fig2}
\end{figure*} 
%

We compute subhalo luminosities as a function of radial position and
mass as follows.  We first determine the minimum and maximum of the
radial distribution (typically $r_{\rm min} \sim {\rm kpc}$, $r_{\rm max} \sim 250 {\rm kpc}$) of subhalos as well as the minimum and maximum
subhalo mass at the present epoch (typically $M_{\rm min} \sim 10^4 M_\odot$, $M_{\rm max} \sim 10^{11} M_\odot$) . We then divide the radial
distribution of subhalos into $N_r$ bins.  Each radial bin is then
subdivided into $N_m$ mass bins of equal logarithmic size in mass.  
We then fit the {\it distribution} of luminosities for different values
of $N_r$ and $N_m$ until the maximum deviation of the fit is less than
10\% of the true value. We find that $N_m=N_r=50$ with at least 200
subhalos per bin has errors of at most 10\% of the true value.


The luminosity of a subhalo is a function of its mass as well as its
position within the host halo.  We find that the subhalo luminosity
PDF is well fit by a log-normal distribution, as
\begin{equation}
\label{eq:PofL} P[\ln L_{M,r}]= \frac{1}{\sqrt{2 \pi}}
\frac{1}{\sigma_{M,r}}  \exp \left[ - \frac{[ \ln L_{M,r} - \langle \ln
L_{M,r} \rangle ]^2 }{2 \sigma_{M,r}^2}\right]
\end{equation}
where,
\begin{equation} \langle \ln L_{M,r} \rangle = a_1+ a_2 \ln \left(
\frac{M}{10^5M_\odot} \right) +a_3 \ln \left( \frac{r} {{\rm 50 kpc}}
\right) ,
\label{eq:lum}
\end{equation}
and,
\begin{equation} \sigma_{M,r} = b_1 + b_2 \ln \left(
\frac{M}{10^5 M_\odot} \right) +b_3 \ln \left( \frac{r} {{\rm 50 kpc}}
\right) .
\label{eq:sigma}
\end{equation}
Here, we implicitly assume that the luminosity $L_{M,r}$ is expressed
in units of photons per second.

Assuming the fiducial particle physics parameters as shown in
Eq.~\ref{eq:luminosity}, the best fit parameters for the whole
population of subhalos in all 200 realizations are $a_1=77.5$,
$a_2=0.87$, $a_3=-0.22$, $b_1=0.75$, $b_2=-0.0026$, and $b_3=0.0061$
(see ``All" in Table~\ref{table:fitparam}). This result can be 
scaled to any assumed particle physics parameters, by simply
adding the term
\begin{equation} 
\ln \left( \frac{ N_\gamma^{\rm tot} \langle \sigma v \rangle M_\chi^{-2}}{9 \times 10^{-29} {\rm cm}^3 {\rm s}^{-1} {\rm GeV}^{-2}} \right) 
\end{equation}
to the parameter $a_1$ of Eq.~\ref{eq:lum}

The fitting function [Eq.~\ref{eq:PofL} with Eq.~(\ref{eq:lum}) \& Eq.~(\ref{eq:sigma})] 
is good to within $\sim$~a few~$\%$ for $P[\ln L_{M,r}]$ as a function of $\ln
L_{M,r}$ over the range of masses and radii we have examined with
sufficient statistics, $M \approx[10^4 - 10^{10}] M_\odot$, and
$r\approx [ 5 - 250]\, {\rm kpc}$.  The mean and variance of the
distribution are functions of mass and radius, reflecting the fact
that the PDF of luminosity is set by the interplay between the mass
function of accreted objects, the redshift of accretion, and the
orbital evolution of the individual subhalos constituting the
population.

\begin{table}
\begin{center}
\begin{tabular}{c|c|c|c|c|c|c}
 & $\ \ a_1 \ \ $& $\ \ a_2 \ \ $ & $\ \ a_3\ \ $  & $\ \ b_1\ \ $ & $\ \ b_2\ \ $ & $\ \ b_3\ \ $\\
\hline
{\rm All}      & 77.4 & 0.87 & \ -0.22\  & 0.75 & \ -0.0026\  & \ 0.0061\  \\
$ {\rm C }_0 $ & 77.4 & 0.87 & \ -0.23\ & 0.74 & -0.0030 & \ 0.011\  \\
$ {\rm C}_+$   & 77.5 & 0.87 & \ -0.26\ & 0.76 & -0.0021 & \  0.0077\  \\
$ {\rm C}_-$   & 77.3 & 0.87 & \ -0.18\ & 0.75 & -0.0013 & \ 0.0043\  \\
\end{tabular} 
\caption{Fitting parameters for the mean and width of the $\gamma$-ray
  annihilation flux distribution function (see text).  }
\label{table:fitparam}
\end{center}
\end{table}

The mean luminosity of isolated, field halos that have {\it not}
experienced strong interactions within the potential of a more massive
halo, scales as $L \propto \rho_s^2 r_s^3 \propto M c^3 / f^2(c)$,
where $M$ is the halo mass, corresponding to a virial radius radius $R
\propto M^{1/3}$, $c \equiv R / r_s$ is the concentration and $f(c)
\propto c^{0.4}$ near $c \approx 30$ as is relevant for small subhalos
\cite{Koushiappas:2003bn}.  If we assume a weak dependence of concentration on mass 
$c(M) \propto M^{-0.1}$ \cite{Bullock:1999he,Wechsler:2001cs,Neto:2007vq,Maccio':2006nu,Klypin:2010qw},
the luminosity of a halo scales roughly as $L \propto M^{0.8}$.
However, subhalo populations deviate from this scaling somewhat for
several reasons.  Subhalos merge into the host halo at a variety of
times, so they sample the $c(M)$ relation at a variety of redshifts and
subhalos of fixed mass at the time of merger exhibit a variety of
bound masses at $z=0$ as a result of their distinct orbital evolution
histories.  Therefore, subhalo luminosities at fixed mass are
influenced by the mass and redshift dependence of concentrations and
subhalo orbital properties.  The effects of these physical changes in
the structure and description of subhalos compared to isolated halos
are reflected in the best fit parameters of Eq.~\ref{eq:lum}.  In
particular, we find that the luminosity of {\it evolved} subhalos
scales approximately as $L \propto M^{0.87}$.

%
\begin{figure*}[t!]
\includegraphics[height=5.9cm]{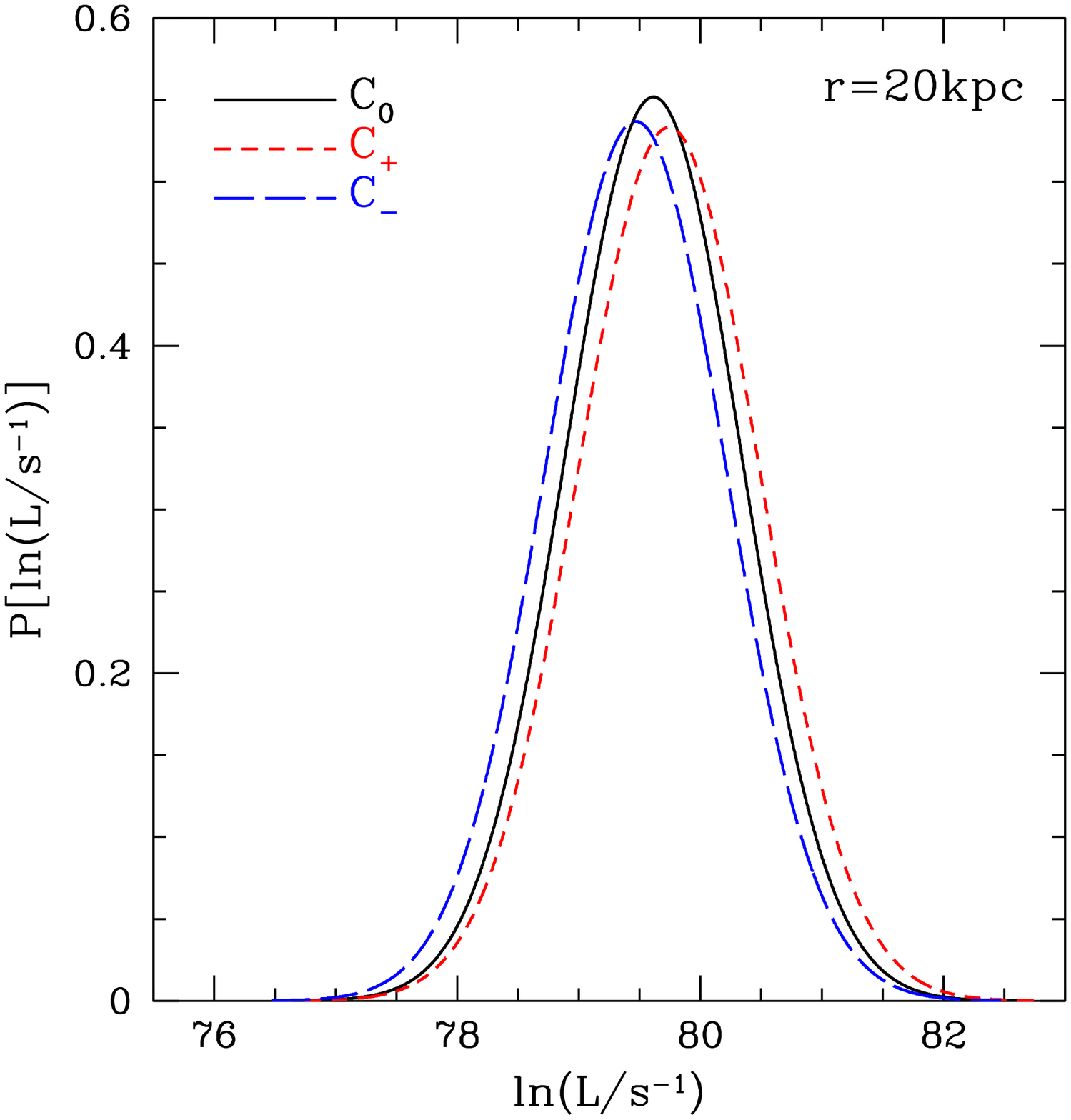}
\includegraphics[height=5.9cm]{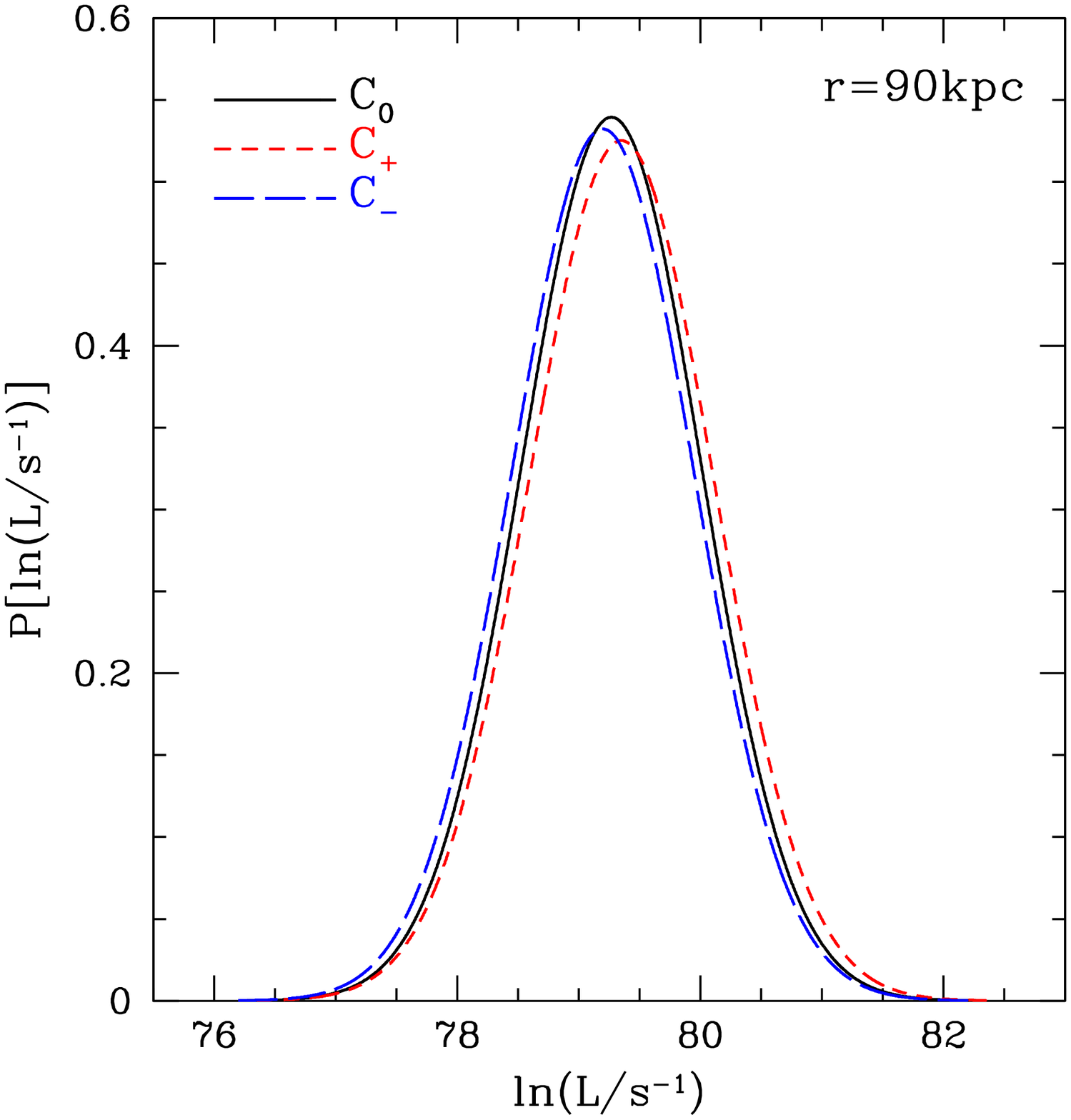}
\includegraphics[height=5.9cm]{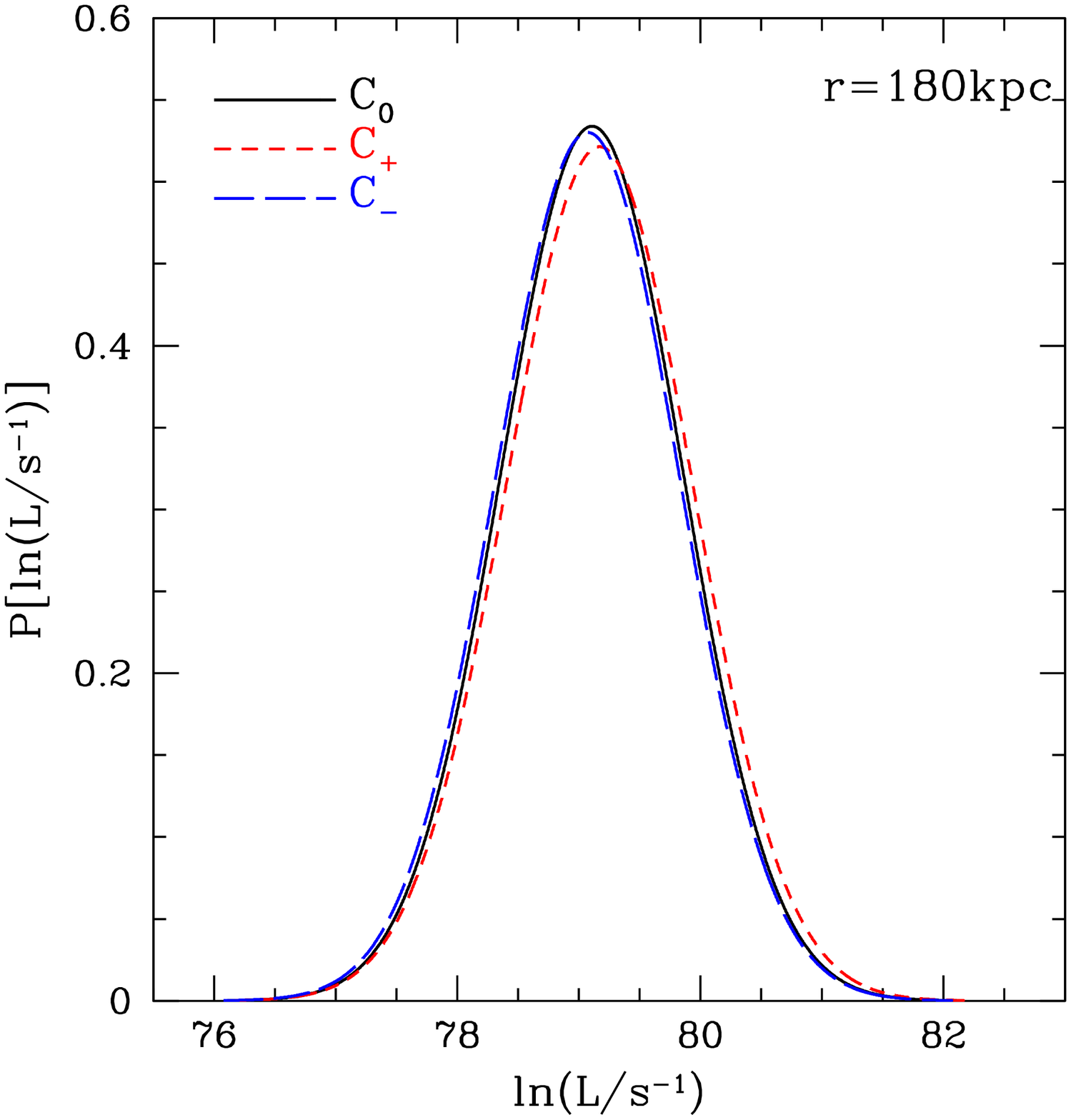}
\caption {Probability distribution functions of the $\gamma$-ray
luminosity of a $10^6 M_\odot$ subhalo at three different radial
distances from the Galactic center (Eq.~\ref{eq:PofL} and parameters
from Table~\ref{table:fitparam}). The solid curve depicts a histogram
of the distribution of luminosities in host halos with a concentration
in the 68 percentile about the mean of the sample of 200 realizations
($C_0$). The short-dashed curve shows the subhalo luminosity PDF for
host halos with a concentration in the upper 16 percent ($C_+$), while
the long-dashed curve shows the subhalo luminosity PDF for host halos
with a concentration in the lower 16 percent ($C_-$). }
\label{fig:fig3}
\end{figure*} 

In Figure~\ref{fig:fig2} we show the distribution of luminosities in
three examples of radial and mass bins.  The figure also shows the fit
to the luminosity distribution given in Eq.~(\ref{eq:PofL}).  The
choice of mass and radius displayed in Fig.~\ref{fig:fig2} are only
meant to demonstrate schematically the agreement between the fitting
functions and the numerical results.  Notice that the spread in
luminosities at fixed mass and position can be roughly {\em an order of
magnitude}, and that the peak of the luminosity PDF depends on 
both, mass as well as radius.

As the structural properties and abundances of subhalos are influenced
by the merger history and properties of the host halo, we expect that
the luminosity PDF will be affected by the distribution of dark matter
within the host dark matter halo. 
In order to explore the relationship between the host halo properties
and the subhalo luminosity PDF we divided the subhalo populations into
three groups based upon the concentrations of their host halos.  We
grouped subhalos with host halo concentrations within the 68\% range
of the mean concentration, $\bar{c}=9.7$.  This group, ``$C_0$",
consists of subhalos in primaries with concentrations in the range
$6.7 \le c \le 13.4$.  We then collected all subhalos in hosts in the
upper and lower $16\%$ ranges, with group $C_-$ consisting of host
concentrations in the range $c < 6.7$ and group $C_+$ consisting of
primaries with $c > 13.4$.  The fitting parameters of
Eq.~(\ref{eq:PofL}) for these three concentration bins are also given
in Table~\ref{table:fitparam}.

Fig.~\ref{fig:fig3} shows the effects of the host halo concentration
on the luminosity PDF of subhalos. At inner radii (left panel of
Fig.~\ref{fig:fig3}), the effects are more pronounced. This is the
region where earliest merging subhalos reside. Host halos with high
concentrations are formed early, and contain subhalos that on average
have formed earlier, and so are also more concentrated
\cite{Wechsler:2001cs}. As the annihilation signal is sensitive to the
concentration of subhalos \cite{Koushiappas:2003bn}, the luminosity of
subhalos in a high concentration host is slightly higher ($C_+$ curve
in Fig.~\ref{fig:fig3}) than subhalos residing in host halos with
lower concentration ($C_-$ curve in Fig.~\ref{fig:fig3}).  At large
radii, $r > r_s$ (right panel of Fig.~\ref{fig:fig3}) the difference
between host halos of different concentrations diminishes, reflecting
the fact that the outer regions of halos typically contain
recently-merged substructure (the host concentration influences subhalo
dynamics little beyond the halo scale radius).  Fig.~\ref{fig:fig3} and Table~\ref{table:fitparam} 
demonstrate that the maximum shift in typical subhalo luminosities is in the inner
regions of host halos, and is relatively small, $\sim 20\%$.

\section{Applications}
\label{section:app}

The flux PDFs that we provide have numerous potential applications.
As an example, in this section we compute the probability distribution
for measuring a flux between $F$ and $F+dF$ from {\em any individual subhalo}
along a line of sight at an angle $\psi$, from the Galactic center.
We compute this ``single halo flux PDF" as
\begin{eqnarray} P_1[F | \psi] &\propto& 
\int_0^{{\ell_{\rm max}}} \int_{\Mmin}^{\Mmax}  \ell^4 \,
\frac{dN[r(\ell,\psi)]}{dMdV} \nonumber \\
&\times& \, P[L_{F,\ell}|M,r(\ell,\psi)]  \, dM \, d \ell
\label{eq:PofF}
\end{eqnarray}
Here, $\ell$ is the line-of-sight distance, $dN[r(\ell,\psi)]/dMdV$ is
the subhalo mass function, and $L_{F,\ell} = 4 \pi \ell^2 F$ ensures a
proper flux measurement for a subhalo of luminosity $L$ at a distance
$\ell$.  The quantity $\ell_{\rm max}$ is the maximum line-of-sight
distance we consider, given by
\begin{equation} 
\ell_{\rm max} = d_\odot \left[ \cos \psi + \sqrt{ (R_{\rm G} / d_\odot)^2 - \sin^2 \psi}\right],
\end{equation}
where $d_\odot = 8 \, {\rm kpc}$ is the distance of the Sun to the
Galactic center, and $R_{\rm G} =250 \, {\rm kpc}$ is the approximate
radius of the Galactic halo.  As in Sec.~\ref{section:importance}, we
assume a mass function of the form $dN/dMdV \propto M^{-\alpha} /
\tilde{r} ( 1 + \tilde{r})$, with $\alpha = 1.9$, as predicted by both
N-body simulations \cite{Springel:2008cc}, as well as by the
semi-analytic model of subhalo populations we use for this study.
Both simulations and our analytic method show little evidence that
$\alpha$ varies significantly as a function of radius
\cite{Zentner:2004dq,Diemand:2006ik,Springel:2008cc}, so any variations are subtle (though 
they may depend upon global host halo properties).  Consequently, 
it is convenient and informative to couple standard halo mass functions 
with our luminosity PDFs to estimate relevant observable quantities.  
However, we emphasize that this particular choice of mass function is
not unique and, in principle, the flux PDF can be derived using
Eq.~(\ref{eq:PofL}) with a variant subhalo mass function.

\begin{figure}[t!]
\includegraphics[height=7.8cm]{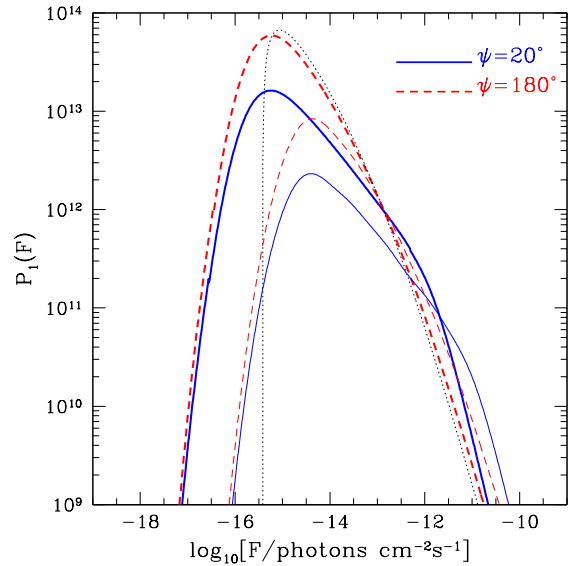}
\caption {The flux probability distribution function derived using
Eq.~\ref{eq:PofL} and the $C_0$ parameters of
Table~\ref{table:fitparam}. The solid blue lines is the flux PDF along
a line of sight at $20^\circ$ relative to the Galactic center, while
the dashed red lines are along $180^\circ$. Thick lines correspond to
a minimum subhalo mass of $10^4 M_\odot$, while thin lines correspond
to a minimum subhalo mass of $10^5 M_\odot$. The black dotted line depicts the
single halo flux PDF derived under the assumption of a delta function
luminosity PDF. 
}
\label{fig:fig4}
\end{figure}

In Figure~\ref{fig:fig4} we show the normalized single halo flux
probability distribution function for two different lines of sight
generated from the contributions of subhalos in the range 
$M_{\rm min} = [ 10^4 - 10^{10}] M_\odot$ (thick lines) and $M_{\rm min} = [ 10^5 -
10^{10}] M_\odot$ (thin lines). This choice of
subhalo mass does not affect the overall shape of the mass function. A
change in the value of $M_{\rm max}$ does not affect the result as the
mass function power law is a steep decreasing function of mass (see
also the left panel of Fig.~\ref{fig:fig0}, which shows how the average flux
due to high-mass halos is lower than that due to low-mass halos).  
The high-flux power-law shape of the flux PDF can be understood in the following
way. As the mass function is proportional to $dN/dMdV \propto
M^{-1.9}$, and the luminosity of a subhalo scales with mass as $L
\propto M^{0.87}$ (see Table I), then the integrand is proportional to
$L^{-2.03}$.  However, $F \propto L/\ell^2$, so the flux PDF is a power
law with a shape given roughly by $P[F] \propto F^{-2.03}$.  This is
apparent for both lines of sight in the high-flux regime.  

Changes in the value of $M_{\rm min}$ affect the low-flux behavior of the flux
PDF. It is easier to understand the low-flux cutoff if we assume a delta
function luminosity PDF instead of the log-normal distribution, as in
\cite{Lee:2008fm}. For a line of sight at some angle $\psi$, there is
a maximum distance $\ell_{\rm max}$ which is a function of $\psi$ and
the radius of the Milky Way halo. If we assume that $L \propto M$, and
that the subhalo radial distribution is the same for all masses, then
the {\it minimum} flux would be given by $F_{\rm min} \propto M_{\rm min}
/ \ell_{\rm max}^2$, i.e., the smaller the minimum mass, the broader
the $P_1(F)$. For illustrative purposes we show the $P_1(F)$ flux PDF
derived under the assumption of a delta function luminosity PDF as in
\cite{Lee:2008fm} in Fig.~\ref{fig:fig4}. Assuming the log-normal
luminosity PDF (instead of a delta function), results in a tail in the
low-flux region of the $P_1(F)$, and thus a broader $P_1(F)$ PDF.

The slope of the flux PDF for low fluxes does depend on the angle
between the line of sight and the Galactic center (see the blue solid,
and red dashed lines in Fig.~\ref{fig:fig4}). At small angles, there
is flux probability excess in the low flux regime relative to the high
flux region. This can be explained using Fig.~\ref{fig:fig0}.  At
large angles from the Galactic center, the total number of halos
intercepted along a line of sight is smaller than the total number of
halos intercepted along a line of sight that passes near the Galactic
center. In addition, the fraction of these halos that are close to the
Sun is smaller for large $\psi$. Therefore, any effects due to the
spread of luminosities will be more pronounced where the fraction of
contributing sources is large, and also nearby (due to the $1/\ell^2$
term). The spread of luminosities thus introduces a spread in flux,
giving rise to substantial a change in the power-law behavior of the flux PDF.

It is important to draw the distinction between the single halo flux
PDF (Eq.~\ref{eq:PofF}) and the probability distribution function of
measuring a {\it total} flux $P(F)$, from the contribution of numerous
halos along the line of sight.  The total flux PDF can be computed in
a straightforward manner from the basic quantity $P_1(F)$
\cite{Scheuer:1957,Barcons:1992ApJ,1990MNRAS.243..366B,1982ApSS..86....3F,2000MNRAS.319..591W},
and more recently \cite{Lee:2008fm}). A thorough investigation of the
total flux PDF using the single halo PDF $P_1(F)$, is presented in
\cite{Baxter:2010}, where the authors evaluate the ability of FGST to
discover dark matter via $\gamma$-rays from Galactic substructure.

Another possible application of the luminosity PDF is in the intrinsic
spread of the expected $\gamma$-ray luminosity signal from dwarf
spheroidal galaxies in the Milky Way halo. Dwarf spheroidals are very
low surface brightness systems dynamically bound to the dark matter
halo of the Milky Way. Their very high mass-to-light ratios make them
ideal for $\gamma$-ray studies as their dark matter distributions can
be well constrained using the velocity dispersions of their stellar
populations. Numerous studies addressed the possibility of detecting
dark matter annihilation in dwarf spheroidals using either FGST or
ground-based \v{C}erenkov telescopes, such as VERITAS and
H.E.S.S. \cite{Baltz:1999ra,Tyler:2002ux,Evans:2003sc,Profumo:2005xd,Bergstrom:2005qk,Strigari:2007at}.
Dynamical studies of the velocity dispersion of stars in these systems
constrain the dark matter mass (and profile)
\cite{Strigari:2006rd,Strigari:2007at}. Suppose that a group of
dwarf spheroidal galaxies have masses estimated via dynamical
measurements. The luminosity PDF presented in
Sec.~\ref{section:pdf_results} can be used to assess the expected
variance of a possible detection of $\gamma$-rays from either
\v{C}erenkov telescopes or the FGST. The estimated variance can be
used in concert with the measurement of the distribution of dark
matter from dynamical studies in the interpretation of any signals 
or limits from these systems.

Moreover, the use of the luminosity PDF can be important in studies
aimed at disentangling the different source contributions to the
diffuse gamma-ray background, such as blazars, starburst galaxies,
pulsars, supernova remnants, and potentially cataclysmic binary
systems, and dark matter
\cite{Ando:2006cr,Ando:2006mt,Ando:2005xg,Miniati:2007ke,Cuoco:2006tr,Cuoco:2007sh,
Taoso:2008qz,Fornasa:2009qh,SiegalGaskins:2008ge,Zhang:2004tj,Pavlidou:2002va,Hooper:2007be}.
A useful tool in these studies is the use of the 
angular correlation function of flux fluctuations, which is estimated
from the convolution of the emissivity as a function of distance and
the spatial correlation function of the sources.  Knowledge of the
expected distribution in luminosities of contributing dark matter
substructure can be used to remove the dark matter contamination to
the astrophysical background, and thus potentially open the window for
the detection of $\gamma$-rays from yet undiscovered sources
\cite{Miniati:2007ke,Pinzke:2010st,Pavlidou:2006rb}.

Finally, the annihilation luminosity PDF can be used in studies aimed
at estimating the likelihood that nearby dark matter subhalos
contribute significantly to the measured antimatter flux
\cite{Adriani:2008zr,Boezio:2009zz}. A dark matter explanation for the
anomalous excess in antimatter flux must rely on either non-standard
extensions to the standard model of particle physics 
(prominent annihilation to charged leptons 
\cite{Cholis:2008hb,Zurek:2008qg,Fox:2008kb,Chen:2008dh}, 
or some new, long-range force \cite{ArkaniHamed:2008qn}, or potentially the
presence of a dark matter halo in the near solar-system neighborhood
\cite{Hooper:2008kv}.  The luminosity PDF presented in this paper can
be used to assess the minimum spread of the expected flux from a given
subhalo and the likelihoods of particular flux measurements from rare, 
nearby objects.  Note however, that propagation effects as well as the
degeneracy between luminosity and the square of the distance to the
subhalo, and velocity-dependent annihilation 
\cite{Robertson:2009bh,Kuhlen:2009kx,Cline:2010ag} 
can all increase the spread considerably.

\section{Conclusions}
\label{sec:conclusions}

We have presented an estimate of the luminosity probability
distribution function of dark matter subhalos within a Milky Way-like
parent halo.  An empirical fit to our numerical calculations suggest
that the luminosity PDF can be described well by a log-normal
distribution with subhalo mass- and position-dependent mean and
variance. This log-normal probability distribution has a width that is
determined by the wide distribution of formation times and
concentrations for both the host halo and subhalos.

The derived luminosity PDF can be used as an ingredient in a number of
interesting calculations regarding predictions for observable dark
matter annihilation products.  This distillation of a complex set of
halo properties should be particularly useful with the impending data
from the Fermi Gamma-ray Space Telescope and continued advances of
ground-based Air Cerenkov Telescopes as well as neutrino telescopes
and antimatter detection instruments.  The tool we provide can be 
used to address such observations, including variance in the
predicted signal, in a relatively simple manner.

As a straightforward example of the application of the PDF, we have estimated
the distribution of observed $\gamma$-ray fluxes along lines of sight as a
function of the angular separation between the line-of-sight and the
Galactic center.  This may be used as a signature to diagnose
unresolved annihilation in a population of Galactic subhalos (see \cite{Baxter:2010}, where the authors 
utilize this distribution to estimate the robustness of
estimating dark matter properties from the diffuse flux measurements of FGST).  
Additional applications of the
annihilation luminosity PDF include estimates in the spread of the
angular power spectrum of flux fluctuations as a probe of unresolved
substructure in the Milky Way, as well as the antimatter flux
distribution from nearby subhalos. These applications make the luminosity PDF a useful tool in the analysis of forthcoming data in the ongoing effort to
identify the dark matter.


\begin{acknowledgments}
SMK and AVK thank the Center for Scientific Computation and
Mathematical Modeling at the University of Maryland College Park for
hospitality. We thank Eric J. Baxter, Scott Dodelson and Louie Strigari for 
useful comments. ARZ acknowledges the support and hospitality of the
Michigan Center for Theoretical Physics at the University of
Michigan. SMK is funded by the NSF and by Brown University. 
ARZ is funded by the University of Pittsburgh, by the NSF 
through grant AST-0806367, and by the DoE.  
AVK is supported by the DoE and NSF grant
AST-0708154, and by the Kavli Institute for
Cosmological Physics at the University of Chicago through the NSF
grant PHY-0551142 and an endowment from the Kavli Foundation. 
\end{acknowledgments}


\bibliography{manuscript}

\begin{thebibliography}{87}
\expandafter\ifx\csname natexlab\endcsname\relax\def\natexlab#1{#1}\fi
\expandafter\ifx\csname bibnamefont\endcsname\relax
  \def\bibnamefont#1{#1}\fi
\expandafter\ifx\csname bibfnamefont\endcsname\relax
  \def\bibfnamefont#1{#1}\fi
\expandafter\ifx\csname citenamefont\endcsname\relax
  \def\citenamefont#1{#1}\fi
\expandafter\ifx\csname url\endcsname\relax
  \def\url#1{\texttt{#1}}\fi
\expandafter\ifx\csname urlprefix\endcsname\relax\def\urlprefix{URL }\fi
\providecommand{\bibinfo}[2]{#2}
\providecommand{\eprint}[2][]{\url{#2}}

\bibitem[{\citenamefont{Jungman et~al.}(1996)\citenamefont{Jungman,
  Kamionkowski, and Griest}}]{Jungman:1995df}
\bibinfo{author}{\bibfnamefont{G.}~\bibnamefont{Jungman}},
  \bibinfo{author}{\bibfnamefont{M.}~\bibnamefont{Kamionkowski}},
  \bibnamefont{and} \bibinfo{author}{\bibfnamefont{K.}~\bibnamefont{Griest}},
  \bibinfo{journal}{Phys. Rept.} \textbf{\bibinfo{volume}{267}},
  \bibinfo{pages}{195} (\bibinfo{year}{1996}).

\bibitem[{\citenamefont{Bertone et~al.}(2005)}]{BHS05}
\bibinfo{author}{\bibfnamefont{G.}~\bibnamefont{Bertone}} \bibnamefont{et~al.},
  \bibinfo{journal}{Phys. Rept.} \textbf{\bibinfo{volume}{405}},
  \bibinfo{pages}{279} (\bibinfo{year}{2005}).

\bibitem[{\citenamefont{Hooper and Profumo}(2007)}]{Hooper:2007qk}
\bibinfo{author}{\bibfnamefont{D.}~\bibnamefont{Hooper}} \bibnamefont{and}
  \bibinfo{author}{\bibfnamefont{S.}~\bibnamefont{Profumo}},
  \bibinfo{journal}{Phys. Rept.} \textbf{\bibinfo{volume}{453}},
  \bibinfo{pages}{29} (\bibinfo{year}{2007}), \eprint{hep-ph/0701197}.

\bibitem[{\citenamefont{Komatsu et~al.}(2009)}]{Komatsu:2008hk}
\bibinfo{author}{\bibfnamefont{E.}~\bibnamefont{Komatsu}} \bibnamefont{et~al.}
  (\bibinfo{collaboration}{WMAP}), \bibinfo{journal}{Astrophys. J. Suppl.}
  \textbf{\bibinfo{volume}{180}}, \bibinfo{pages}{330} (\bibinfo{year}{2009}),
  \eprint{0803.0547}.

\bibitem[{\citenamefont{Klypin et~al.}(2002)\citenamefont{Klypin, Zhao, and
  Somerville}}]{Klypin:2001xu}
\bibinfo{author}{\bibfnamefont{A.}~\bibnamefont{Klypin}},
  \bibinfo{author}{\bibfnamefont{H.}~\bibnamefont{Zhao}}, \bibnamefont{and}
  \bibinfo{author}{\bibfnamefont{R.~S.} \bibnamefont{Somerville}},
  \bibinfo{journal}{Astrophys. J.} \textbf{\bibinfo{volume}{573}},
  \bibinfo{pages}{597} (\bibinfo{year}{2002}), \eprint{astro-ph/0110390}.

\bibitem[{\citenamefont{Li and White}(2008)}]{Li:2007eg}
\bibinfo{author}{\bibfnamefont{Y.-S.} \bibnamefont{Li}} \bibnamefont{and}
  \bibinfo{author}{\bibfnamefont{S.~D.~M.} \bibnamefont{White}},
  \bibinfo{journal}{Mon. Not. Roy. Astron. Soc.}
  \textbf{\bibinfo{volume}{384}}, \bibinfo{pages}{1459} (\bibinfo{year}{2008}),
  \eprint{0710.3740}.

\bibitem[{\citenamefont{Schmid et~al.}(1999)}]{Schmid:1998mx}
\bibinfo{author}{\bibfnamefont{C.}~\bibnamefont{Schmid}} \bibnamefont{et~al.},
  \bibinfo{journal}{Phys. Rev.} \textbf{\bibinfo{volume}{D59}},
  \bibinfo{pages}{043517} (\bibinfo{year}{1999}).

\bibitem[{\citenamefont{Green et~al.}(2004)\citenamefont{Green, Hofmann, and
  Schwarz}}]{Green:2003un}
\bibinfo{author}{\bibfnamefont{A.~M.} \bibnamefont{Green}},
  \bibinfo{author}{\bibfnamefont{S.}~\bibnamefont{Hofmann}}, \bibnamefont{and}
  \bibinfo{author}{\bibfnamefont{D.~J.} \bibnamefont{Schwarz}},
  \bibinfo{journal}{Mon. Not. Roy. Astron. Soc.}
  \textbf{\bibinfo{volume}{353}}, \bibinfo{pages}{L23} (\bibinfo{year}{2004}).

\bibitem[{\citenamefont{Green et~al.}(2005)}]{Green:2005fa}
\bibinfo{author}{\bibfnamefont{A.~M.} \bibnamefont{Green}}
  \bibnamefont{et~al.}, \bibinfo{journal}{JCAP}
  \textbf{\bibinfo{volume}{0508}}, \bibinfo{pages}{003} (\bibinfo{year}{2005}).

\bibitem[{\citenamefont{Hofmann et~al.}(2001)}]{HSS01}
\bibinfo{author}{\bibfnamefont{S.}~\bibnamefont{Hofmann}} \bibnamefont{et~al.},
  \bibinfo{journal}{Phys. Rev.} \textbf{\bibinfo{volume}{D64}},
  \bibinfo{pages}{083507} (\bibinfo{year}{2001}).

\bibitem[{\citenamefont{Dodelson et~al.}(2008)\citenamefont{Dodelson, Hooper,
  and Serpico}}]{Dodelson:2007gd}
\bibinfo{author}{\bibfnamefont{S.}~\bibnamefont{Dodelson}},
  \bibinfo{author}{\bibfnamefont{D.}~\bibnamefont{Hooper}}, \bibnamefont{and}
  \bibinfo{author}{\bibfnamefont{P.~D.} \bibnamefont{Serpico}},
  \bibinfo{journal}{Phys. Rev.} \textbf{\bibinfo{volume}{D77}},
  \bibinfo{pages}{063512} (\bibinfo{year}{2008}), \eprint{0711.4621}.

\bibitem[{\citenamefont{Serpico and Zaharijas}(2008)}]{Serpico:2008ga}
\bibinfo{author}{\bibfnamefont{P.~D.} \bibnamefont{Serpico}} \bibnamefont{and}
  \bibinfo{author}{\bibfnamefont{G.}~\bibnamefont{Zaharijas}},
  \bibinfo{journal}{Astropart. Phys.} \textbf{\bibinfo{volume}{29}},
  \bibinfo{pages}{380} (\bibinfo{year}{2008}), \eprint{0802.3245}.

\bibitem[{\citenamefont{Gondolo and Silk}(1999)}]{Gondolo:1999ef}
\bibinfo{author}{\bibfnamefont{P.}~\bibnamefont{Gondolo}} \bibnamefont{and}
  \bibinfo{author}{\bibfnamefont{J.}~\bibnamefont{Silk}},
  \bibinfo{journal}{Phys. Rev. Lett.} \textbf{\bibinfo{volume}{83}},
  \bibinfo{pages}{1719} (\bibinfo{year}{1999}), \eprint{astro-ph/9906391}.

\bibitem[{\citenamefont{Horns}(2005)}]{Horns:2004bk}
\bibinfo{author}{\bibfnamefont{D.}~\bibnamefont{Horns}},
  \bibinfo{journal}{Phys. Lett.} \textbf{\bibinfo{volume}{B607}},
  \bibinfo{pages}{225} (\bibinfo{year}{2005}), \eprint{astro-ph/0408192}.

\bibitem[{\citenamefont{Merritt et~al.}(2002)\citenamefont{Merritt,
  Milosavljevic, Verde, and Jimenez}}]{Merritt:2002vj}
\bibinfo{author}{\bibfnamefont{D.}~\bibnamefont{Merritt}},
  \bibinfo{author}{\bibfnamefont{M.}~\bibnamefont{Milosavljevic}},
  \bibinfo{author}{\bibfnamefont{L.}~\bibnamefont{Verde}}, \bibnamefont{and}
  \bibinfo{author}{\bibfnamefont{R.}~\bibnamefont{Jimenez}},
  \bibinfo{journal}{Phys. Rev. Lett.} \textbf{\bibinfo{volume}{88}},
  \bibinfo{pages}{191301} (\bibinfo{year}{2002}), \eprint{astro-ph/0201376}.

\bibitem[{\citenamefont{Bergstrom et~al.}(1998)\citenamefont{Bergstrom, Ullio,
  and Buckley}}]{Bergstrom:1997fj}
\bibinfo{author}{\bibfnamefont{L.}~\bibnamefont{Bergstrom}},
  \bibinfo{author}{\bibfnamefont{P.}~\bibnamefont{Ullio}}, \bibnamefont{and}
  \bibinfo{author}{\bibfnamefont{J.~H.} \bibnamefont{Buckley}},
  \bibinfo{journal}{Astropart. Phys.} \textbf{\bibinfo{volume}{9}},
  \bibinfo{pages}{137} (\bibinfo{year}{1998}), \eprint{astro-ph/9712318}.

\bibitem[{\citenamefont{Aloisio et~al.}(2004)\citenamefont{Aloisio, Blasi, and
  Olinto}}]{Aloisio:2004hy}
\bibinfo{author}{\bibfnamefont{R.}~\bibnamefont{Aloisio}},
  \bibinfo{author}{\bibfnamefont{P.}~\bibnamefont{Blasi}}, \bibnamefont{and}
  \bibinfo{author}{\bibfnamefont{A.~V.} \bibnamefont{Olinto}},
  \bibinfo{journal}{JCAP} \textbf{\bibinfo{volume}{0405}}, \bibinfo{pages}{007}
  (\bibinfo{year}{2004}), \eprint{astro-ph/0402588}.

\bibitem[{\citenamefont{Zaharijas and Hooper}(2006)}]{Zaharijas:2006qb}
\bibinfo{author}{\bibfnamefont{G.}~\bibnamefont{Zaharijas}} \bibnamefont{and}
  \bibinfo{author}{\bibfnamefont{D.}~\bibnamefont{Hooper}},
  \bibinfo{journal}{Phys. Rev.} \textbf{\bibinfo{volume}{D73}},
  \bibinfo{pages}{103501} (\bibinfo{year}{2006}), \eprint{astro-ph/0603540}.

\bibitem[{\citenamefont{Baltz et~al.}(2000)\citenamefont{Baltz, Briot, Salati,
  Taillet, and Silk}}]{Baltz:1999ra}
\bibinfo{author}{\bibfnamefont{E.~A.} \bibnamefont{Baltz}},
  \bibinfo{author}{\bibfnamefont{C.}~\bibnamefont{Briot}},
  \bibinfo{author}{\bibfnamefont{P.}~\bibnamefont{Salati}},
  \bibinfo{author}{\bibfnamefont{R.}~\bibnamefont{Taillet}}, \bibnamefont{and}
  \bibinfo{author}{\bibfnamefont{J.}~\bibnamefont{Silk}},
  \bibinfo{journal}{Phys. Rev.} \textbf{\bibinfo{volume}{D61}},
  \bibinfo{pages}{023514} (\bibinfo{year}{2000}), \eprint{astro-ph/9909112}.

\bibitem[{\citenamefont{Tyler}(2002)}]{Tyler:2002ux}
\bibinfo{author}{\bibfnamefont{C.}~\bibnamefont{Tyler}},
  \bibinfo{journal}{Phys. Rev.} \textbf{\bibinfo{volume}{D66}},
  \bibinfo{pages}{023509} (\bibinfo{year}{2002}), \eprint{astro-ph/0203242}.

\bibitem[{\citenamefont{Evans et~al.}(2004)\citenamefont{Evans, Ferrer, and
  Sarkar}}]{Evans:2003sc}
\bibinfo{author}{\bibfnamefont{N.~W.} \bibnamefont{Evans}},
  \bibinfo{author}{\bibfnamefont{F.}~\bibnamefont{Ferrer}}, \bibnamefont{and}
  \bibinfo{author}{\bibfnamefont{S.}~\bibnamefont{Sarkar}},
  \bibinfo{journal}{Phys. Rev.} \textbf{\bibinfo{volume}{D69}},
  \bibinfo{pages}{123501} (\bibinfo{year}{2004}), \eprint{astro-ph/0311145}.

\bibitem[{\citenamefont{Profumo}(2005)}]{Profumo:2005xd}
\bibinfo{author}{\bibfnamefont{S.}~\bibnamefont{Profumo}},
  \bibinfo{journal}{Phys. Rev.} \textbf{\bibinfo{volume}{D72}},
  \bibinfo{pages}{103521} (\bibinfo{year}{2005}), \eprint{astro-ph/0508628}.

\bibitem[{\citenamefont{Bergstrom and Hooper}(2006)}]{Bergstrom:2005qk}
\bibinfo{author}{\bibfnamefont{L.}~\bibnamefont{Bergstrom}} \bibnamefont{and}
  \bibinfo{author}{\bibfnamefont{D.}~\bibnamefont{Hooper}},
  \bibinfo{journal}{Phys. Rev.} \textbf{\bibinfo{volume}{D73}},
  \bibinfo{pages}{063510} (\bibinfo{year}{2006}), \eprint{hep-ph/0512317}.

\bibitem[{\citenamefont{Calcaneo-Roldan and
  Moore}(2000)}]{Calcaneo-Roldan:2000yt}
\bibinfo{author}{\bibfnamefont{C.}~\bibnamefont{Calcaneo-Roldan}}
  \bibnamefont{and} \bibinfo{author}{\bibfnamefont{B.}~\bibnamefont{Moore}},
  \bibinfo{journal}{Phys. Rev.} \textbf{\bibinfo{volume}{D62}},
  \bibinfo{pages}{123005} (\bibinfo{year}{2000}), \eprint{astro-ph/0010056}.

\bibitem[{\citenamefont{Tasitsiomi and Olinto}(2002)}]{Tasitsiomi:2002vh}
\bibinfo{author}{\bibfnamefont{A.}~\bibnamefont{Tasitsiomi}} \bibnamefont{and}
  \bibinfo{author}{\bibfnamefont{A.~V.} \bibnamefont{Olinto}},
  \bibinfo{journal}{Phys. Rev.} \textbf{\bibinfo{volume}{D66}},
  \bibinfo{pages}{083006} (\bibinfo{year}{2002}), \eprint{astro-ph/0206040}.

\bibitem[{\citenamefont{Stoehr et~al.}(2003)\citenamefont{Stoehr, White,
  Springel, Tormen, and Yoshida}}]{Stoehr:2003hf}
\bibinfo{author}{\bibfnamefont{F.}~\bibnamefont{Stoehr}},
  \bibinfo{author}{\bibfnamefont{S.~D.~M.} \bibnamefont{White}},
  \bibinfo{author}{\bibfnamefont{V.}~\bibnamefont{Springel}},
  \bibinfo{author}{\bibfnamefont{G.}~\bibnamefont{Tormen}}, \bibnamefont{and}
  \bibinfo{author}{\bibfnamefont{N.}~\bibnamefont{Yoshida}},
  \bibinfo{journal}{Mon. Not. Roy. Astron. Soc.}
  \textbf{\bibinfo{volume}{345}}, \bibinfo{pages}{1313} (\bibinfo{year}{2003}),
  \eprint{astro-ph/0307026}.

\bibitem[{\citenamefont{Koushiappas et~al.}(2004)\citenamefont{Koushiappas,
  Zentner, and Walker}}]{Koushiappas:2003bn}
\bibinfo{author}{\bibfnamefont{S.~M.} \bibnamefont{Koushiappas}},
  \bibinfo{author}{\bibfnamefont{A.~R.} \bibnamefont{Zentner}},
  \bibnamefont{and} \bibinfo{author}{\bibfnamefont{T.~P.}
  \bibnamefont{Walker}}, \bibinfo{journal}{Phys. Rev.}
  \textbf{\bibinfo{volume}{D69}}, \bibinfo{pages}{043501}
  (\bibinfo{year}{2004}), \eprint{astro-ph/0309464}.

\bibitem[{\citenamefont{Baltz et~al.}(2007)\citenamefont{Baltz, Taylor, and
  Wai}}]{Baltz:2006sv}
\bibinfo{author}{\bibfnamefont{E.~A.} \bibnamefont{Baltz}},
  \bibinfo{author}{\bibfnamefont{J.~E.} \bibnamefont{Taylor}},
  \bibnamefont{and} \bibinfo{author}{\bibfnamefont{L.~L.} \bibnamefont{Wai}},
  \bibinfo{journal}{Astrophys. J. Lett.} \textbf{\bibinfo{volume}{659}},
  \bibinfo{pages}{L125} (\bibinfo{year}{2007}), \eprint{astro-ph/0610731}.

\bibitem[{\citenamefont{Strigari
  et~al.}(2007{\natexlab{a}})\citenamefont{Strigari, Koushiappas, Bullock, and
  Kaplinghat}}]{Strigari:2006rd}
\bibinfo{author}{\bibfnamefont{L.~E.} \bibnamefont{Strigari}},
  \bibinfo{author}{\bibfnamefont{S.~M.} \bibnamefont{Koushiappas}},
  \bibinfo{author}{\bibfnamefont{J.~S.} \bibnamefont{Bullock}},
  \bibnamefont{and}
  \bibinfo{author}{\bibfnamefont{M.}~\bibnamefont{Kaplinghat}},
  \bibinfo{journal}{Phys. Rev.} \textbf{\bibinfo{volume}{D75}},
  \bibinfo{pages}{083526} (\bibinfo{year}{2007}{\natexlab{a}}),
  \eprint{astro-ph/0611925}.

\bibitem[{\citenamefont{Diemand et~al.}(2007)\citenamefont{Diemand, Kuhlen, and
  Madau}}]{Diemand:2006ik}
\bibinfo{author}{\bibfnamefont{J.}~\bibnamefont{Diemand}},
  \bibinfo{author}{\bibfnamefont{M.}~\bibnamefont{Kuhlen}}, \bibnamefont{and}
  \bibinfo{author}{\bibfnamefont{P.}~\bibnamefont{Madau}},
  \bibinfo{journal}{Astrophys. J.} \textbf{\bibinfo{volume}{657}},
  \bibinfo{pages}{262} (\bibinfo{year}{2007}), \eprint{astro-ph/0611370}.

\bibitem[{\citenamefont{Strigari et~al.}(2007{\natexlab{b}})}]{Strigari:2007at}
\bibinfo{author}{\bibfnamefont{L.~E.} \bibnamefont{Strigari}}
  \bibnamefont{et~al.} (\bibinfo{year}{2007}{\natexlab{b}}),
  \eprint{0709.1510}.

\bibitem[{\citenamefont{Bringmann}(2009)}]{Bringmann:2009vf}
\bibinfo{author}{\bibfnamefont{T.}~\bibnamefont{Bringmann}},
  \bibinfo{journal}{New J. Phys.} \textbf{\bibinfo{volume}{11}},
  \bibinfo{pages}{105027} (\bibinfo{year}{2009}), \eprint{0903.0189}.

\bibitem[{\citenamefont{Koushiappas}(2009)}]{Koushiappas:2009du}
\bibinfo{author}{\bibfnamefont{S.~M.} \bibnamefont{Koushiappas}},
  \bibinfo{journal}{New J. Phys.} \textbf{\bibinfo{volume}{11}},
  \bibinfo{pages}{105012} (\bibinfo{year}{2009}), \eprint{0905.1998}.

\bibitem[{\citenamefont{Bringmann and Hofmann}(2007)}]{Bringmann:2006mu}
\bibinfo{author}{\bibfnamefont{T.}~\bibnamefont{Bringmann}} \bibnamefont{and}
  \bibinfo{author}{\bibfnamefont{S.}~\bibnamefont{Hofmann}},
  \bibinfo{journal}{JCAP} \textbf{\bibinfo{volume}{0407}}, \bibinfo{pages}{016}
  (\bibinfo{year}{2007}), \eprint{hep-ph/0612238}.

\bibitem[{\citenamefont{Pieri et~al.}(2005)}]{Pieri:2005pg}
\bibinfo{author}{\bibfnamefont{L.}~\bibnamefont{Pieri}} \bibnamefont{et~al.},
  \bibinfo{journal}{Phys. Rev. Lett.} \textbf{\bibinfo{volume}{95}},
  \bibinfo{pages}{211301} (\bibinfo{year}{2005}).

\bibitem[{\citenamefont{Ando et~al.}(2008)\citenamefont{Ando, Kamionkowski,
  Lee, and Koushiappas}}]{Ando:2008br}
\bibinfo{author}{\bibfnamefont{S.}~\bibnamefont{Ando}},
  \bibinfo{author}{\bibfnamefont{M.}~\bibnamefont{Kamionkowski}},
  \bibinfo{author}{\bibfnamefont{S.~K.} \bibnamefont{Lee}}, \bibnamefont{and}
  \bibinfo{author}{\bibfnamefont{S.~M.} \bibnamefont{Koushiappas}},
  \bibinfo{journal}{Phys. Rev.} \textbf{\bibinfo{volume}{D78}},
  \bibinfo{pages}{101301} (\bibinfo{year}{2008}), \eprint{0809.0886}.

\bibitem[{\citenamefont{Ando}(2009)}]{Ando:2009fp}
\bibinfo{author}{\bibfnamefont{S.}~\bibnamefont{Ando}}, \bibinfo{journal}{Phys.
  Rev.} \textbf{\bibinfo{volume}{D80}}, \bibinfo{pages}{023520}
  (\bibinfo{year}{2009}), \eprint{0903.4685}.

\bibitem[{\citenamefont{Ando et~al.}(2007{\natexlab{a}})\citenamefont{Ando,
  Komatsu, Narumoto, and Totani}}]{Ando:2006cr}
\bibinfo{author}{\bibfnamefont{S.}~\bibnamefont{Ando}},
  \bibinfo{author}{\bibfnamefont{E.}~\bibnamefont{Komatsu}},
  \bibinfo{author}{\bibfnamefont{T.}~\bibnamefont{Narumoto}}, \bibnamefont{and}
  \bibinfo{author}{\bibfnamefont{T.}~\bibnamefont{Totani}},
  \bibinfo{journal}{Phys. Rev.} \textbf{\bibinfo{volume}{D75}},
  \bibinfo{pages}{063519} (\bibinfo{year}{2007}{\natexlab{a}}),
  \eprint{astro-ph/0612467}.

\bibitem[{\citenamefont{Cuoco et~al.}(2007)}]{Cuoco:2006tr}
\bibinfo{author}{\bibfnamefont{A.}~\bibnamefont{Cuoco}} \bibnamefont{et~al.},
  \bibinfo{journal}{JCAP} \textbf{\bibinfo{volume}{0704}}, \bibinfo{pages}{013}
  (\bibinfo{year}{2007}), \eprint{astro-ph/0612559}.

\bibitem[{\citenamefont{Cuoco et~al.}(2008)\citenamefont{Cuoco, Brandbyge,
  Hannestad, Haugboelle, and Miele}}]{Cuoco:2007sh}
\bibinfo{author}{\bibfnamefont{A.}~\bibnamefont{Cuoco}},
  \bibinfo{author}{\bibfnamefont{J.}~\bibnamefont{Brandbyge}},
  \bibinfo{author}{\bibfnamefont{S.}~\bibnamefont{Hannestad}},
  \bibinfo{author}{\bibfnamefont{T.}~\bibnamefont{Haugboelle}},
  \bibnamefont{and} \bibinfo{author}{\bibfnamefont{G.}~\bibnamefont{Miele}},
  \bibinfo{journal}{Phys. Rev.} \textbf{\bibinfo{volume}{D77}},
  \bibinfo{pages}{123518} (\bibinfo{year}{2008}), \eprint{0710.4136}.

\bibitem[{\citenamefont{Fornasa et~al.}(2009)\citenamefont{Fornasa, Pieri,
  Bertone, and Branchini}}]{Fornasa:2009qh}
\bibinfo{author}{\bibfnamefont{M.}~\bibnamefont{Fornasa}},
  \bibinfo{author}{\bibfnamefont{L.}~\bibnamefont{Pieri}},
  \bibinfo{author}{\bibfnamefont{G.}~\bibnamefont{Bertone}}, \bibnamefont{and}
  \bibinfo{author}{\bibfnamefont{E.}~\bibnamefont{Branchini}},
  \bibinfo{journal}{Phys. Rev.} \textbf{\bibinfo{volume}{D80}},
  \bibinfo{pages}{023518} (\bibinfo{year}{2009}), \eprint{0901.2921}.

\bibitem[{\citenamefont{Hooper and Serpico}(2007)}]{Hooper:2007be}
\bibinfo{author}{\bibfnamefont{D.}~\bibnamefont{Hooper}} \bibnamefont{and}
  \bibinfo{author}{\bibfnamefont{P.~D.} \bibnamefont{Serpico}},
  \bibinfo{journal}{JCAP} \textbf{\bibinfo{volume}{0706}}, \bibinfo{pages}{013}
  (\bibinfo{year}{2007}), \eprint{astro-ph/0702328}.

\bibitem[{\citenamefont{Lee et~al.}(2009)\citenamefont{Lee, Ando, and
  Kamionkowski}}]{Lee:2008fm}
\bibinfo{author}{\bibfnamefont{S.~K.} \bibnamefont{Lee}},
  \bibinfo{author}{\bibfnamefont{S.}~\bibnamefont{Ando}}, \bibnamefont{and}
  \bibinfo{author}{\bibfnamefont{M.}~\bibnamefont{Kamionkowski}},
  \bibinfo{journal}{JCAP} \textbf{\bibinfo{volume}{0907}}, \bibinfo{pages}{007}
  (\bibinfo{year}{2009}), \eprint{0810.1284}.

\bibitem[{\citenamefont{Siegal-Gaskins}(2009)}]{SiegalGaskins:2009pz}
\bibinfo{author}{\bibfnamefont{J.~M.} \bibnamefont{Siegal-Gaskins}}
  (\bibinfo{year}{2009}), \eprint{0907.0183}.

\bibitem[{\citenamefont{Siegal-Gaskins and
  Pavlidou}(2009)}]{SiegalGaskins:2009ux}
\bibinfo{author}{\bibfnamefont{J.~M.} \bibnamefont{Siegal-Gaskins}}
  \bibnamefont{and} \bibinfo{author}{\bibfnamefont{V.}~\bibnamefont{Pavlidou}},
  \bibinfo{journal}{Phys. Rev. Lett.} \textbf{\bibinfo{volume}{102}},
  \bibinfo{pages}{241301} (\bibinfo{year}{2009}), \eprint{0901.3776}.

\bibitem[{\citenamefont{Siegal-Gaskins}(2008)}]{SiegalGaskins:2008ge}
\bibinfo{author}{\bibfnamefont{J.~M.} \bibnamefont{Siegal-Gaskins}},
  \bibinfo{journal}{JCAP} \textbf{\bibinfo{volume}{0810}}, \bibinfo{pages}{040}
  (\bibinfo{year}{2008}), \eprint{0807.1328}.

\bibitem[{\citenamefont{Taoso et~al.}(2009)\citenamefont{Taoso, Ando, Bertone,
  and Profumo}}]{Taoso:2008qz}
\bibinfo{author}{\bibfnamefont{M.}~\bibnamefont{Taoso}},
  \bibinfo{author}{\bibfnamefont{S.}~\bibnamefont{Ando}},
  \bibinfo{author}{\bibfnamefont{G.}~\bibnamefont{Bertone}}, \bibnamefont{and}
  \bibinfo{author}{\bibfnamefont{S.}~\bibnamefont{Profumo}},
  \bibinfo{journal}{Phys. Rev.} \textbf{\bibinfo{volume}{D79}},
  \bibinfo{pages}{043521} (\bibinfo{year}{2009}), \eprint{0811.4493}.

\bibitem[{\citenamefont{Zentner et~al.}(2005)\citenamefont{Zentner, Berlind,
  Bullock, Kravtsov, and Wechsler}}]{Zentner:2004dq}
\bibinfo{author}{\bibfnamefont{A.~R.} \bibnamefont{Zentner}},
  \bibinfo{author}{\bibfnamefont{A.~A.} \bibnamefont{Berlind}},
  \bibinfo{author}{\bibfnamefont{J.~S.} \bibnamefont{Bullock}},
  \bibinfo{author}{\bibfnamefont{A.~V.} \bibnamefont{Kravtsov}},
  \bibnamefont{and} \bibinfo{author}{\bibfnamefont{R.~H.}
  \bibnamefont{Wechsler}}, \bibinfo{journal}{Astrophys. J.}
  \textbf{\bibinfo{volume}{624}}, \bibinfo{pages}{505} (\bibinfo{year}{2005}),
  \eprint{astro-ph/0411586}.

\bibitem[{\citenamefont{Bullock et~al.}(2001)}]{Bullock:1999he}
\bibinfo{author}{\bibfnamefont{J.~S.} \bibnamefont{Bullock}}
  \bibnamefont{et~al.}, \bibinfo{journal}{Mon. Not. Roy. Astron. Soc.}
  \textbf{\bibinfo{volume}{321}}, \bibinfo{pages}{559} (\bibinfo{year}{2001}).

\bibitem[{\citenamefont{Wechsler et~al.}(2002)\citenamefont{Wechsler, Bullock,
  Primack, Kravtsov, and Dekel}}]{Wechsler:2001cs}
\bibinfo{author}{\bibfnamefont{R.~H.} \bibnamefont{Wechsler}},
  \bibinfo{author}{\bibfnamefont{J.~S.} \bibnamefont{Bullock}},
  \bibinfo{author}{\bibfnamefont{J.~R.} \bibnamefont{Primack}},
  \bibinfo{author}{\bibfnamefont{A.~V.} \bibnamefont{Kravtsov}},
  \bibnamefont{and} \bibinfo{author}{\bibfnamefont{A.}~\bibnamefont{Dekel}},
  \bibinfo{journal}{Astrophys. J.} \textbf{\bibinfo{volume}{568}},
  \bibinfo{pages}{52} (\bibinfo{year}{2002}), \eprint{astro-ph/0108151}.

\bibitem[{\citenamefont{Ando et~al.}(2007{\natexlab{b}})\citenamefont{Ando,
  Komatsu, Narumoto, and Totani}}]{Ando:2006mt}
\bibinfo{author}{\bibfnamefont{S.}~\bibnamefont{Ando}},
  \bibinfo{author}{\bibfnamefont{E.}~\bibnamefont{Komatsu}},
  \bibinfo{author}{\bibfnamefont{T.}~\bibnamefont{Narumoto}}, \bibnamefont{and}
  \bibinfo{author}{\bibfnamefont{T.}~\bibnamefont{Totani}},
  \bibinfo{journal}{Mon. Not. Roy. Astron. Soc.}
  \textbf{\bibinfo{volume}{376}}, \bibinfo{pages}{1635}
  (\bibinfo{year}{2007}{\natexlab{b}}), \eprint{astro-ph/0610155}.

\bibitem[{\citenamefont{Miniati et~al.}(2007)\citenamefont{Miniati,
  Koushiappas, and Di~Matteo}}]{Miniati:2007ke}
\bibinfo{author}{\bibfnamefont{F.}~\bibnamefont{Miniati}},
  \bibinfo{author}{\bibfnamefont{S.~M.} \bibnamefont{Koushiappas}},
  \bibnamefont{and}
  \bibinfo{author}{\bibfnamefont{T.}~\bibnamefont{Di~Matteo}},
  \bibinfo{journal}{Astrophys. J.} \textbf{\bibinfo{volume}{667}},
  \bibinfo{pages}{L1} (\bibinfo{year}{2007}), \eprint{astro-ph/0702083}.

\bibitem[{\citenamefont{Zhang and Beacom}(2004)}]{Zhang:2004tj}
\bibinfo{author}{\bibfnamefont{P.-J.} \bibnamefont{Zhang}} \bibnamefont{and}
  \bibinfo{author}{\bibfnamefont{J.~F.} \bibnamefont{Beacom}},
  \bibinfo{journal}{Astrophys. J.} \textbf{\bibinfo{volume}{614}},
  \bibinfo{pages}{37} (\bibinfo{year}{2004}), \eprint{astro-ph/0401351}.

\bibitem[{\citenamefont{Pavlidou and Fields}(2002)}]{Pavlidou:2002va}
\bibinfo{author}{\bibfnamefont{V.}~\bibnamefont{Pavlidou}} \bibnamefont{and}
  \bibinfo{author}{\bibfnamefont{B.~D.} \bibnamefont{Fields}},
  \bibinfo{journal}{Astrophys. J.} \textbf{\bibinfo{volume}{575}},
  \bibinfo{pages}{L5} (\bibinfo{year}{2002}), \eprint{astro-ph/0207253}.

\bibitem[{\citenamefont{Hooper et~al.}(2009)\citenamefont{Hooper, Stebbins, and
  Zurek}}]{Hooper:2008kv}
\bibinfo{author}{\bibfnamefont{D.}~\bibnamefont{Hooper}},
  \bibinfo{author}{\bibfnamefont{A.}~\bibnamefont{Stebbins}}, \bibnamefont{and}
  \bibinfo{author}{\bibfnamefont{K.~M.} \bibnamefont{Zurek}},
  \bibinfo{journal}{Phys. Rev.} \textbf{\bibinfo{volume}{D79}},
  \bibinfo{pages}{103513} (\bibinfo{year}{2009}), \eprint{0812.3202}.

\bibitem[{\citenamefont{Kuhlen et~al.}(2008{\natexlab{a}})\citenamefont{Kuhlen,
  Diemand, and Madau}}]{kuhlen:2008aw}
\bibinfo{author}{\bibfnamefont{M.}~\bibnamefont{Kuhlen}},
  \bibinfo{author}{\bibfnamefont{J.}~\bibnamefont{Diemand}}, \bibnamefont{and}
  \bibinfo{author}{\bibfnamefont{P.}~\bibnamefont{Madau}}
  (\bibinfo{year}{2008}{\natexlab{a}}), \eprint{0805.4416}.

\bibitem[{\citenamefont{{Navarro} et~al.}(1996)\citenamefont{{Navarro},
  {Frenk}, and {White}}}]{NFW96}
\bibinfo{author}{\bibfnamefont{J.~F.} \bibnamefont{{Navarro}}},
  \bibinfo{author}{\bibfnamefont{C.~S.} \bibnamefont{{Frenk}}},
  \bibnamefont{and} \bibinfo{author}{\bibfnamefont{S.~D.~M.}
  \bibnamefont{{White}}}, \bibinfo{journal}{\apj}
  \textbf{\bibinfo{volume}{462}}, \bibinfo{pages}{563} (\bibinfo{year}{1996}),
  \eprint{arXiv:astro-ph/9508025}.

\bibitem[{\citenamefont{Navarro et~al.}(1997)\citenamefont{Navarro, Frenk, and
  White}}]{Navarro:1996gj}
\bibinfo{author}{\bibfnamefont{J.~F.} \bibnamefont{Navarro}},
  \bibinfo{author}{\bibfnamefont{C.~S.} \bibnamefont{Frenk}}, \bibnamefont{and}
  \bibinfo{author}{\bibfnamefont{S.~D.~M.} \bibnamefont{White}},
  \bibinfo{journal}{Astrophys. J.} \textbf{\bibinfo{volume}{490}},
  \bibinfo{pages}{493} (\bibinfo{year}{1997}), \eprint{astro-ph/9611107}.

\bibitem[{\citenamefont{D'Onghia et~al.}(2010)\citenamefont{D'Onghia, Springel,
  Hernquist, and Keres}}]{D'Onghia:2009pz}
\bibinfo{author}{\bibfnamefont{E.}~\bibnamefont{D'Onghia}},
  \bibinfo{author}{\bibfnamefont{V.}~\bibnamefont{Springel}},
  \bibinfo{author}{\bibfnamefont{L.}~\bibnamefont{Hernquist}},
  \bibnamefont{and} \bibinfo{author}{\bibfnamefont{D.}~\bibnamefont{Keres}},
  \bibinfo{journal}{Astrophys. J.} \textbf{\bibinfo{volume}{709}},
  \bibinfo{pages}{1138} (\bibinfo{year}{2010}), \eprint{0907.3482}.

\bibitem[{\citenamefont{Kuhlen et~al.}(2008{\natexlab{b}})\citenamefont{Kuhlen,
  Diemand, Madau, and Zemp}}]{Kuhlen:2008qj}
\bibinfo{author}{\bibfnamefont{M.}~\bibnamefont{Kuhlen}},
  \bibinfo{author}{\bibfnamefont{J.}~\bibnamefont{Diemand}},
  \bibinfo{author}{\bibfnamefont{P.}~\bibnamefont{Madau}}, \bibnamefont{and}
  \bibinfo{author}{\bibfnamefont{M.}~\bibnamefont{Zemp}}, \bibinfo{journal}{J.
  Phys. Conf. Ser.} \textbf{\bibinfo{volume}{125}}, \bibinfo{pages}{012008}
  (\bibinfo{year}{2008}{\natexlab{b}}), \eprint{0810.3614}.

\bibitem[{\citenamefont{Boylan-Kolchin
  et~al.}(2009)\citenamefont{Boylan-Kolchin, Springel, White, and
  Jenkins}}]{BoylanKolchin:2009an}
\bibinfo{author}{\bibfnamefont{M.}~\bibnamefont{Boylan-Kolchin}},
  \bibinfo{author}{\bibfnamefont{V.}~\bibnamefont{Springel}},
  \bibinfo{author}{\bibfnamefont{S.~D.~M.} \bibnamefont{White}},
  \bibnamefont{and} \bibinfo{author}{\bibfnamefont{A.}~\bibnamefont{Jenkins}}
  (\bibinfo{year}{2009}), \eprint{0911.4484}.

\bibitem[{\citenamefont{Kazantzidis et~al.}(2004)}]{Kazantzidis:2003hb}
\bibinfo{author}{\bibfnamefont{S.}~\bibnamefont{Kazantzidis}}
  \bibnamefont{et~al.}, \bibinfo{journal}{Astrophys. J.}
  \textbf{\bibinfo{volume}{608}}, \bibinfo{pages}{663} (\bibinfo{year}{2004}),
  \eprint{astro-ph/0312194}.

\bibitem[{\citenamefont{Kazantzidis et~al.}(2009)\citenamefont{Kazantzidis,
  Zentner, Kravtsov, Bullock, and Debattista}}]{Kazantzidis:2009zq}
\bibinfo{author}{\bibfnamefont{S.}~\bibnamefont{Kazantzidis}},
  \bibinfo{author}{\bibfnamefont{A.~R.} \bibnamefont{Zentner}},
  \bibinfo{author}{\bibfnamefont{A.~V.} \bibnamefont{Kravtsov}},
  \bibinfo{author}{\bibfnamefont{J.~S.} \bibnamefont{Bullock}},
  \bibnamefont{and} \bibinfo{author}{\bibfnamefont{V.~P.}
  \bibnamefont{Debattista}}, \bibinfo{journal}{Astrophys. J.}
  \textbf{\bibinfo{volume}{700}}, \bibinfo{pages}{1896} (\bibinfo{year}{2009}),
  \eprint{0902.1983}.

\bibitem[{\citenamefont{Reed and Koushiappas}(2010)}]{Reed:2010}
\bibinfo{author}{\bibfnamefont{D.~S.} \bibnamefont{Reed}} \bibnamefont{and}
  \bibinfo{author}{\bibfnamefont{S.~M.} \bibnamefont{Koushiappas}}
  (\bibinfo{year}{2010}), \bibinfo{note}{in preparation}.

\bibitem[{\citenamefont{Robertson and Zentner}(2009)}]{Robertson:2009bh}
\bibinfo{author}{\bibfnamefont{B.}~\bibnamefont{Robertson}} \bibnamefont{and}
  \bibinfo{author}{\bibfnamefont{A.}~\bibnamefont{Zentner}},
  \bibinfo{journal}{Phys. Rev.} \textbf{\bibinfo{volume}{D79}},
  \bibinfo{pages}{083525} (\bibinfo{year}{2009}), \eprint{0902.0362}.

\bibitem[{\citenamefont{Neto et~al.}(2007)}]{Neto:2007vq}
\bibinfo{author}{\bibfnamefont{A.~F.} \bibnamefont{Neto}} \bibnamefont{et~al.}
  (\bibinfo{year}{2007}), \eprint{0706.2919}.

\bibitem[{\citenamefont{Maccio' et~al.}(2007)}]{Maccio':2006nu}
\bibinfo{author}{\bibfnamefont{A.~V.} \bibnamefont{Maccio'}}
  \bibnamefont{et~al.}, \bibinfo{journal}{Mon. Not. Roy. Astron. Soc.}
  \textbf{\bibinfo{volume}{378}}, \bibinfo{pages}{55} (\bibinfo{year}{2007}),
  \eprint{astro-ph/0608157}.

\bibitem[{\citenamefont{Klypin et~al.}(2010)\citenamefont{Klypin,
  Trujillo-Gomez, and Primack}}]{Klypin:2010qw}
\bibinfo{author}{\bibfnamefont{A.}~\bibnamefont{Klypin}},
  \bibinfo{author}{\bibfnamefont{S.}~\bibnamefont{Trujillo-Gomez}},
  \bibnamefont{and} \bibinfo{author}{\bibfnamefont{J.}~\bibnamefont{Primack}}
  (\bibinfo{year}{2010}), \eprint{1002.3660}.

\bibitem[{\citenamefont{Springel et~al.}(2008)}]{Springel:2008cc}
\bibinfo{author}{\bibfnamefont{V.}~\bibnamefont{Springel}}
  \bibnamefont{et~al.}, \bibinfo{journal}{Mon. Not. Roy. Astron. Soc.}
  \textbf{\bibinfo{volume}{391}}, \bibinfo{pages}{1685} (\bibinfo{year}{2008}),
  \eprint{0809.0898}.

\bibitem[{\citenamefont{{Barcons}}(1992)}]{Barcons:1992ApJ}
\bibinfo{author}{\bibfnamefont{X.}~\bibnamefont{{Barcons}}},
  \bibinfo{journal}{Astrophys. J.} \textbf{\bibinfo{volume}{396}},
  \bibinfo{pages}{460} (\bibinfo{year}{1992}).

\bibitem[{\citenamefont{Scheuer and Ryle}(1957)}]{Scheuer:1957}
\bibinfo{author}{\bibfnamefont{P.~A.~G.} \bibnamefont{Scheuer}}
  \bibnamefont{and} \bibinfo{author}{\bibfnamefont{M.}~\bibnamefont{Ryle}},
  \bibinfo{journal}{Proc. Camb. Phil. Soc.} \textbf{\bibinfo{volume}{53}},
  \bibinfo{pages}{764} (\bibinfo{year}{1957}).

\bibitem[{\citenamefont{Barcons and Fabian}(1990)}]{1990MNRAS.243..366B}
\bibinfo{author}{\bibfnamefont{X.}~\bibnamefont{Barcons}} \bibnamefont{and}
  \bibinfo{author}{\bibfnamefont{A.~C.} \bibnamefont{Fabian}},
  \bibinfo{journal}{Mon. Not. R. Astron. Soc.} \textbf{\bibinfo{volume}{243}},
  \bibinfo{pages}{366} (\bibinfo{year}{1990}).

\bibitem[{\citenamefont{{Franceschini}}(1982)}]{1982ApSS..86....3F}
\bibinfo{author}{\bibfnamefont{A.}~\bibnamefont{{Franceschini}}},
  \bibinfo{journal}{Astrophys. J. Supp.} \textbf{\bibinfo{volume}{86}},
  \bibinfo{pages}{3} (\bibinfo{year}{1982}).

\bibitem[{\citenamefont{Windridge and Phillipps}(2000)}]{2000MNRAS.319..591W}
\bibinfo{author}{\bibfnamefont{D.}~\bibnamefont{Windridge}} \bibnamefont{and}
  \bibinfo{author}{\bibfnamefont{S.}~\bibnamefont{Phillipps}},
  \bibinfo{journal}{Mon. Not. R. Astron. Soc.} \textbf{\bibinfo{volume}{319}},
  \bibinfo{pages}{591} (\bibinfo{year}{2000}).

\bibitem[{\citenamefont{Baxter et~al.}(2010)\citenamefont{Baxter, Dodelson,
  Koushiappas, and Strigari}}]{Baxter:2010}
\bibinfo{author}{\bibfnamefont{E.~J.} \bibnamefont{Baxter}},
  \bibinfo{author}{\bibfnamefont{S.}~\bibnamefont{Dodelson}},
  \bibinfo{author}{\bibfnamefont{S.~M.} \bibnamefont{Koushiappas}},
  \bibnamefont{and} \bibinfo{author}{\bibfnamefont{L.~E.}
  \bibnamefont{Strigari}} (\bibinfo{year}{2010}), \eprint{1006.2399}.

\bibitem[{\citenamefont{Ando and Komatsu}(2006)}]{Ando:2005xg}
\bibinfo{author}{\bibfnamefont{S.}~\bibnamefont{Ando}} \bibnamefont{and}
  \bibinfo{author}{\bibfnamefont{E.}~\bibnamefont{Komatsu}},
  \bibinfo{journal}{Phys. Rev.} \textbf{\bibinfo{volume}{D73}},
  \bibinfo{pages}{023521} (\bibinfo{year}{2006}), \eprint{astro-ph/0512217}.

\bibitem[{\citenamefont{Pinzke and Pfrommer}(2010)}]{Pinzke:2010st}
\bibinfo{author}{\bibfnamefont{A.}~\bibnamefont{Pinzke}} \bibnamefont{and}
  \bibinfo{author}{\bibfnamefont{C.}~\bibnamefont{Pfrommer}}
  (\bibinfo{year}{2010}), \eprint{1001.5023}.

\bibitem[{\citenamefont{Pavlidou and Fields}(2006)}]{Pavlidou:2006rb}
\bibinfo{author}{\bibfnamefont{V.}~\bibnamefont{Pavlidou}} \bibnamefont{and}
  \bibinfo{author}{\bibfnamefont{B.~D.} \bibnamefont{Fields}},
  \bibinfo{journal}{Astrophys. J.} \textbf{\bibinfo{volume}{642}},
  \bibinfo{pages}{734} (\bibinfo{year}{2006}), \eprint{astro-ph/0611923}.

\bibitem[{\citenamefont{Adriani et~al.}(2009)}]{Adriani:2008zr}
\bibinfo{author}{\bibfnamefont{O.}~\bibnamefont{Adriani}} \bibnamefont{et~al.}
  (\bibinfo{collaboration}{PAMELA}), \bibinfo{journal}{Nature}
  \textbf{\bibinfo{volume}{458}}, \bibinfo{pages}{607} (\bibinfo{year}{2009}),
  \eprint{0810.4995}.

\bibitem[{\citenamefont{Boezio et~al.}(2009)}]{Boezio:2009zz}
\bibinfo{author}{\bibfnamefont{M.}~\bibnamefont{Boezio}} \bibnamefont{et~al.},
  \bibinfo{journal}{New J. Phys.} \textbf{\bibinfo{volume}{11}},
  \bibinfo{pages}{105023} (\bibinfo{year}{2009}).

\bibitem[{\citenamefont{Cholis et~al.}(2009)\citenamefont{Cholis, Goodenough,
  Hooper, Simet, and Weiner}}]{Cholis:2008hb}
\bibinfo{author}{\bibfnamefont{I.}~\bibnamefont{Cholis}},
  \bibinfo{author}{\bibfnamefont{L.}~\bibnamefont{Goodenough}},
  \bibinfo{author}{\bibfnamefont{D.}~\bibnamefont{Hooper}},
  \bibinfo{author}{\bibfnamefont{M.}~\bibnamefont{Simet}}, \bibnamefont{and}
  \bibinfo{author}{\bibfnamefont{N.}~\bibnamefont{Weiner}},
  \bibinfo{journal}{Phys. Rev.} \textbf{\bibinfo{volume}{D80}},
  \bibinfo{pages}{123511} (\bibinfo{year}{2009}), \eprint{0809.1683}.

\bibitem[{\citenamefont{Zurek}(2009)}]{Zurek:2008qg}
\bibinfo{author}{\bibfnamefont{K.~M.} \bibnamefont{Zurek}},
  \bibinfo{journal}{Phys. Rev.} \textbf{\bibinfo{volume}{D79}},
  \bibinfo{pages}{115002} (\bibinfo{year}{2009}), \eprint{0811.4429}.

\bibitem[{\citenamefont{Fox and Poppitz}(2009)}]{Fox:2008kb}
\bibinfo{author}{\bibfnamefont{P.~J.} \bibnamefont{Fox}} \bibnamefont{and}
  \bibinfo{author}{\bibfnamefont{E.}~\bibnamefont{Poppitz}},
  \bibinfo{journal}{Phys. Rev.} \textbf{\bibinfo{volume}{D79}},
  \bibinfo{pages}{083528} (\bibinfo{year}{2009}), \eprint{0811.0399}.

\bibitem[{\citenamefont{Chen and Takahashi}(2009)}]{Chen:2008dh}
\bibinfo{author}{\bibfnamefont{C.-R.} \bibnamefont{Chen}} \bibnamefont{and}
  \bibinfo{author}{\bibfnamefont{F.}~\bibnamefont{Takahashi}},
  \bibinfo{journal}{JCAP} \textbf{\bibinfo{volume}{0902}}, \bibinfo{pages}{004}
  (\bibinfo{year}{2009}), \eprint{0810.4110}.

\bibitem[{\citenamefont{Arkani-Hamed et~al.}(2009)\citenamefont{Arkani-Hamed,
  Finkbeiner, Slatyer, and Weiner}}]{ArkaniHamed:2008qn}
\bibinfo{author}{\bibfnamefont{N.}~\bibnamefont{Arkani-Hamed}},
  \bibinfo{author}{\bibfnamefont{D.~P.} \bibnamefont{Finkbeiner}},
  \bibinfo{author}{\bibfnamefont{T.~R.} \bibnamefont{Slatyer}},
  \bibnamefont{and} \bibinfo{author}{\bibfnamefont{N.}~\bibnamefont{Weiner}},
  \bibinfo{journal}{Phys. Rev.} \textbf{\bibinfo{volume}{D79}},
  \bibinfo{pages}{015014} (\bibinfo{year}{2009}), \eprint{0810.0713}.

\bibitem[{\citenamefont{Kuhlen et~al.}(2009)\citenamefont{Kuhlen, Madau, and
  Silk}}]{Kuhlen:2009kx}
\bibinfo{author}{\bibfnamefont{M.}~\bibnamefont{Kuhlen}},
  \bibinfo{author}{\bibfnamefont{P.}~\bibnamefont{Madau}}, \bibnamefont{and}
  \bibinfo{author}{\bibfnamefont{J.}~\bibnamefont{Silk}},
  \bibinfo{journal}{Science} \textbf{\bibinfo{volume}{325}},
  \bibinfo{pages}{970} (\bibinfo{year}{2009}), \eprint{0907.0005}.

\bibitem[{\citenamefont{Cline et~al.}(2010)\citenamefont{Cline, Vincent, and
  Xue}}]{Cline:2010ag}
\bibinfo{author}{\bibfnamefont{J.~M.} \bibnamefont{Cline}},
  \bibinfo{author}{\bibfnamefont{A.~C.} \bibnamefont{Vincent}},
  \bibnamefont{and} \bibinfo{author}{\bibfnamefont{W.}~\bibnamefont{Xue}},
  \bibinfo{journal}{Phys. Rev.} \textbf{\bibinfo{volume}{D81}},
  \bibinfo{pages}{083512} (\bibinfo{year}{2010}), \eprint{1001.5399}.

\end{thebibliography}

\end{document}